\documentclass[fleqn,usenatbib]{mnras}

\usepackage[T1]{fontenc}
\usepackage{ae,aecompl}

\usepackage{graphicx}	
\usepackage{amsmath}	
\usepackage{amssymb}	

\usepackage{color}
\usepackage{ulem}

\title[A NIR variable star survey in the SMC]{A near infrared variable star survey in the Magellanic Clouds: \\ The Small Magellanic Cloud data}

\author[Y. Ita et al.]{Yoshifusa Ita$^{1}$\thanks{E-mail: yita@astr.tohoku.ac.jp}, Noriyuki Matsunaga$^{2}$, Toshihiko Tanab\'{e}$^{3}$, Yoshikazu Nakada$^{2,3}$, \newauthor Daisuke Kato$^{4}$, Takahiro Nagayama$^{5}$, Chie Nagashima$^{6}$, Mikio Kurita$^{7}$,
\newauthor Yasushi Nakajima$^{8}$, Patricia A. Whitelock$^{9,10}$, John W. Menzies$^{10}$, Michael W. Feast$^{9,10}$,
\newauthor Tetsuya Nagata$^{7}$, Motohide Tamura$^{2,11}$, and Hidehiko Nakaya$^{12}$ \\
$^{1}$Astronomical Institute, Graduate School of Science, Tohoku University, 6-3 Aramaki Aoba, Aoba-ku, Sendai, Miyagi 980-8578, Japan\\
$^{2}$Department of Astronomy, School of Science, The University of Tokyo, 7-3-1 Hongo, Bunkyo-ku, Tokyo 113-0033, Japan\\
$^{3}$Institute of Astronomy, School of Science, The University of Tokyo, Mitaka, Tokyo 181-0015, Japan\\
$^{4}$Center for Low Carbon Society Strategy, Japan Science and Technology Agency, Science Plaza(4F), 5-3, Yonban-cho, Chiyoda-ku, Tokyo, 102-8666, Japan\\
$^{5}$Department of Science and Engineering, Kagoshima University, Korimoto, Kagoshima 890-0065, Japan\\
$^{6}$Department of Astrophysics, Nagoya University, Chikusa-ku, Nagoya 464-8602, Japan\\
$^{7}$Department of Astronomy, Kyoto University, Kitashirakawa-Oiwake-cho, Sakyo-ku, Kyoto 606-8502, Japan\\
$^{8}$Hitotsubashi University, 2-1 Naka, Kunitachi, Tokyo 186-8601, Japan\\
$^{9}$Department of Astronomy, University of Cape Town, 7701 Rondebosch, South Africa\\
$^{10}$South African Astronomical Observatory, PO Box 9, 7935 Observatory, South Africa\\
$^{11}$National Astronomical Observatory of Japan, 1-2-1 Osawa, Mitaka, Tokyo 181-8588, Japan\\
$^{12}$Advanced Technology Center, National Astronomical Observatory of Japan, 1-2-1, Osawa, Mitaka, Tokyo 181-8588, Japan
}

\date{Accepted XXX. Received YYY; in original form ZZZ}

\pubyear{2015}

\begin{document}
\label{firstpage}
\pagerange{\pageref{firstpage}--\pageref{lastpage}}
\maketitle

\begin{abstract}A very long term near-infrared variable star survey towards the Large and Small Magellanic Clouds was carried out using the 1.4m InfraRed Survey Facility at the South African Astronomical Observatory. This project was initiated in December 2000 in the LMC, and in July 2001 in the SMC. Since then an area of 3 square degrees along the bar in the LMC and an area of 1 square degree in the central part of the SMC have been repeatedly observed. This survey is ongoing, but results obtained with data taken until December 2017 are reported in this paper. Over more than 15 years we have observed the two survey areas more than one hundred times. This is the first survey that provides near-infrared time-series data with such a long time baseline and on such a large scale. This paper describes the observations in the SMC and publishes a point source photometric catalogue, a variable source catalogue, and time-series data. 
\end{abstract}




\section{Introduction}
The Magellanic Clouds offer a number of advantages compared  with the Milky Way where studies of stars are concerned: (1) their distances are reasonably well known, (2) they are located at relatively high Galactic latitude so interstellar extinction is small, (3) they are less contaminated by foreground objects, (4) their metallicities are different from those of most stars in the solar neighbourhood, (5) certain stars, e.g., Cepheids, Miras and RR Lyrae variables, can be calibrated as distance indicators.

In the 1990s a number of monitoring projects were started that were aimed at detecting microlensing events (OGLE, \citealt*{udalski}; MACHO, \citealt{alcock}; EROS, \citealt{afonso}; MOA, \citealt{bond}). As a by-product of these surveys huge data sets of precise time-series photometry for millions of stars in the Large and Small Magellanic Clouds (LMC and SMC, respectively) were obtained. Some of the survey projects are still ongoing with upgraded observing systems, providing very long-term time-series data (e.g., over 20 years for the OGLE projects). 

None of the surveys cited above operates at wavelengths longer than the $I$-band, and most are at shorter wavelengths. Therefore infrared variable sources such as evolved stars with high mass-loss rates and young stellar objects (YSOs) may have escaped detection. Variability is common among evolved stars with high mass-loss rates, and pulsation is believed to play a key role in the mass-loss process. Studying the variability of YSOs is important for understanding the very early stages of stellar evolution. Up to now a number of infrared photometric surveys have been carried out, some of them covering the Magellanic Clouds (e.g., IRAS, Explanatory Supplement\citealt{iras}; ISO, \citealt{kessler1997}; MSX, \citealt{mill1994}; DENIS, \citealt{epchtein}; 2MASS, \citealt{skrutskie}; IRSF/SIRIUS, \citealt{kato}). Recently, both the Spitzer (\citealt{meixner}, \citealt{gordon}) and the AKARI infrared satellites (\citealt{ita2008}, \citealt{ita2010}, \citealt{kato2012}) mapped the Magellanic Clouds in near- to mid-infrared wavebands. Both Spitzer and AKARI have two epochs of data and \citet{vijh2009} compared the photometry of those epochs and thereby identified a large number of infrared variable stars in the LMC. None of the previous infrared surveys had sufficient epochs of observation to properly characterize the variability. 

Before the aforementioned massive optical time-series photometric surveys there were several shallow but large-scale near-infrared multi-epoch surveys toward the Magellanic Clouds. \citet*{evans1988} made $J, H$ and $K$ observation in the Radcliffe Variable Star Field in the SMC, \cite{hughes1990} repeatedly observed a 6$^\circ$$\times$12$^\circ$ area in the LMC in $I, J, H$ and $K$, and \citet*{reid1995} repeatedly observed the northern part of the LMC in $J, H$ and $K$. In addition to these, there are many other infrared variable star surveys in the Magellanic Clouds. \citet{groenewegen1997} provides one of the most comprehensive reviews of earlier work where he reports that ``IRAS triggered researches" were started in the late eighties using the IRAS data products. Those studies concentrated on the brightest AGB stars and red supergiants in the Magellanic Clouds (e.g., \citealt{whitelock2003}) due to the limited sensitivity of the IRAS satellite.

Previous observational studies of variable stars in the Magellanic Clouds can be summarized as follows: (1) there are many time-series photometric surveys, but all of them are in the optical and are insensitive to infrared variables, (2) there are a number of deep photometric surveys in the near-, mid- and far-infrared, but at very few epochs, making it difficult to study the nature of variable stars, and (3) there are infrared multi-epoch surveys, but they are not deep enough to detect all red giants. There are also IRAS, ISO and MSX triggered studies but they are biased toward bright red supergiants and relatively massive (hence bright) AGB variables.

Here we present a moderately deep, near-infrared time-series survey covering one square degree of the central part of the SMC. The data were obtained between July 2001 and December 2017, and there are at least 115 independent epochs at $J, H,$ and $K_s$. Later papers will discuss the details of various types of variable star and similar papers will deal with the LMC survey. The great strengths of this study are the larger number of epochs and the long time baseline. Hence, the data provide excellent mean magnitudes for known variable stars found, for example, by microlensing surveys, and at a cadence that allows the type of variability to be characterised.

The ongoing VISTA survey of the Magellanic Cloud system (VMC, \citealt{cioni2011}) will ultimately provide a deeper survey over a larger area, but with fewer epochs than the one described here. VMC is the near-infrared ($Y, J,$ and $K_s$) multi-epoch survey across an area of about 170 square degrees over the Magellanic Cloud system. The survey is deep enough to measure accurate mean magnitudes for variable stars that are relatively faint in the near-infrared, such as RR Lyrae stars (e.g., \citealt{muraveva2018}) and Cepheids (e.g., \citealt{ripepi2016}). Their data obtained between November 2009 and August 2013 are now publicly available (VMC-DR4), providing at least 3 epochs of photometric data at $Y$ and $J,$ and at least 12 epochs at $K_s$ for selected fields.

\begin{table}
\centering
\caption{The readout noise and gain of the SIRIUS camera.}
\label{readout}
\begin{tabular}{ccc}
\hline
\multicolumn{1}{c}{Array} & \multicolumn{1}{c}{Readout Noise [e$^-$]} & \multicolumn{1}{c}{Gain [e$^-$/ADU]} \\
\hline
\multicolumn{3}{c}{Messia IV} \\
\hline
$J$ & 43 & 5.7 \\
$H$ & 30 & 5.1 \\
$K_s$ & 45 & 5.5 \\
\hline
\multicolumn{3}{c}{Messia V} \\
\hline
$J$ & 19 & 5.1 \\
$H$ & 25 & 5.0 \\
$K_s$ & 35 & 5.3 \\
\hline
\end{tabular}
\end{table}

\section{Observations}
We have repeatedly observed a total area of 3 square degrees along the LMC bar since December 2000, and a total area of 1 square degree around the SMC centre since July 2001 with instruments and a survey strategy described below.

\subsection{InfraRed Survey Facility and the SIRIUS near-infrared camera}
The InfraRed Survey Facility (IRSF) is situated at the SAAO Sutherland station and is operated as a joint Japanese/South African project. It officially opened and started formal operations in November 2000. The IRSF consists of a dedicated 1.4-m alt-azimuth telescope equipped with a near-infrared camera (the ``Simultaneous three-colour InfraRed Imager for Unbiased Surveys" or SIRIUS) for specialized surveys toward the Magellanic Clouds and the centre of the Milky Way. The SIRIUS imager uses three 1024$\times$1024 HgCdTe arrays to make observations simultaneously in three wavebands $J$(1.25$\mu$m), $H$(1.63$\mu$m) and $K_s$(2.14$\mu$m). These filters are similar to the Mauna Kea Observatories (MKO) near-infrared photometric system (\citealt{tokunaga2002}). The SIRIUS camera has a field of view of about 7.7 arcmin square with a scale of 0.453 arcsec/pixel. 

In the initial phase of SIRIUS camera operations the chip readout and control electronics used the Messia IV system, but it was upgraded to the Messia V system after April 2004 to reduce readout time. The readout noise and gain for each system are tabulated in Table~\ref{readout} (Nagayama, private communication). Photometric errors calculated in this work are estimated by using these values. Another mechanical upgrade to add an imaging polarimetry capability (SIRPOL, \citealt{kandori2006}) was made in February 2013. The SIRPOL is removable if necessary, and it should be noted that data used in this work are all taken without SIRPOL. Due to this upgrade, the pixel field of view of the SIRIUS camera became narrower by about 1\%. This change has been accommodated in the data reduction process by matching the scale to the one before February 2013.  Further details of the instrument can be found in \citet{nagashima} and \citet{nagayama}. 

Two large survey projects were planned from the opening of the IRSF. The first was a deep photometric survey toward the Magellanic Clouds aimed at making a near-infrared (NIR) point source catalogue. The results were published by \citet{kato}. It provides NIR point source catalogues of both Magellanic Clouds that are about two magnitudes deeper and four times finer in spatial resolution than the congeneric NIR point source catalogue of the 2MASS survey. The second major project was the variable star survey towards the Magellanic Clouds. A time-series survey of this type requires a great deal of telescope time over many years. The IRSF provides an ideal facility in which to carry out this ``near-infrared variable star survey in the Magellanic Clouds".

\begin{figure}
\centering
\includegraphics[angle=0,bb=0 0 720 720,scale=0.343]{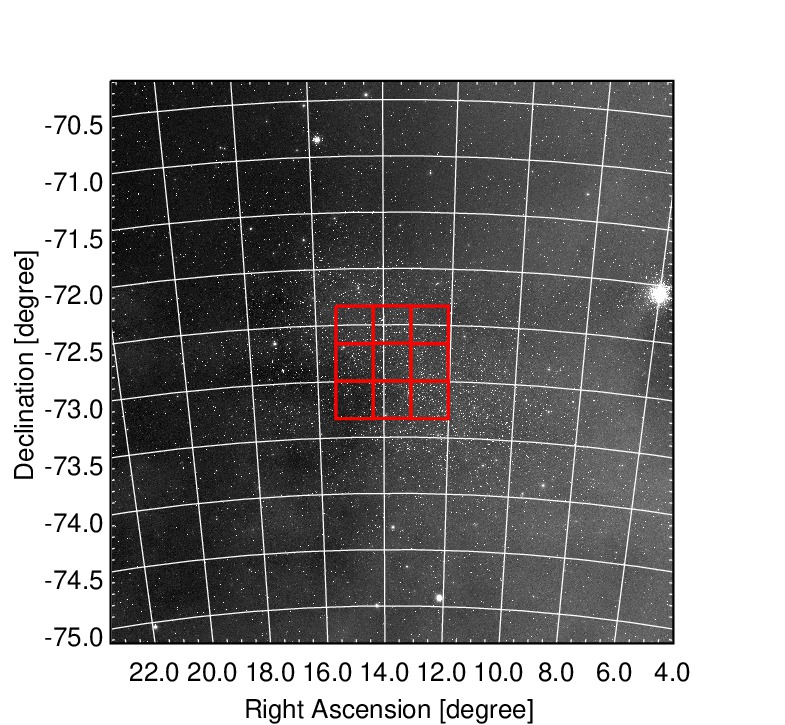}
\caption{A schematic map of the 1$^\circ$$\times$1$^\circ$ survey area in the SMC (thick solid line). Each square mesh has a side of 20 arcmin and consists of nine sub-meshes, which correspond to the field of view of the SIRIUS camera. The background is the 2MASS $J-$band image.}
\label{smc}
\end{figure}

\subsection{Survey strategy and observational specifications}
\label{sec:strategy}
In order to study variable sources in the SMC, an area of 1 square degree around the SMC centre, as shown in Fig.~\ref{smc}, was observed repeatedly. The 1$^\circ$$\times$1$^\circ$ area is divided into nine 20$^\prime$$\times$20$^\prime$ regions. Each of these regions is further subdivided into nine 7$^\prime$$\times$7$^\prime$ fields of view that are labeled from A to I. The central position of each field is given in Appendix A. This survey area was chosen because it is well populated and we can expect the maximum efficiency in gathering a large sample of variable sources. Moreover, OGLE optical microlensing survey data (\citealt{udalski}, \citealt{zebrun2001}) are available for the same area. Their survey and ours are mutually complementary, so the amalgamation of these two data sets will eventually produce a complete variable source catalogue for that area.

The SMC survey commenced in July 2001 and in this paper we present data obtained up to December 2017. During that period, the whole survey area was observed more than 140 times and, after manual removal of poor images (see section \ref{sec:photo}), there are at least 115 (at most 162) independent epochs of photometric data at $J, H,$ and $K_s$.

For the survey, we use a fixed exposure time of 5 seconds. We take ten 5\,s images for each field of view at a time, with a dithering radius of 15 arcsec. This configuration allows us to take data for sources as bright as $\sim$ 9 mag at $J, H$, and $K_s$. The detection limits are derived and described in Section~\ref{sec:photo}. The 5\,s exposure time is chosen so that we can simultaneously detect bright AGB stars as well as red giants well below the tip of the first red giant branch (RGB). Therefore we can measure Cepheid variables and all of the AGB variables \textit{except} extremely red ones (such as those found in NGC419 and NGC1978 by \citet{tanabe1998} and \citet{kamath2010} and those found in the LMC by \citet{gruendl}) and reach several magnitudes below the RGB tip. Extremely red sources and/or faint sources will be detected, and their variability studied, from the data obtained in VMC (\citealt{cioni2011}) and in the SAGE-Var program (e.g., \citealt{riebel2015}), a follow-on to the Spitzer legacy program Surveying the Agents of Galaxy Evolution (SAGE; \citealt{meixner}).

\section{Data Reduction}
All data were reduced (i.e., flat-fielded, dark-current subtracted and sky subtracted) in the same manner using the SIRIUS pipeline software (Nakajima, private communication). A sky image is made from the images acquired with the same sequence before and after taking the images of the subject, after masking bright stars. The SIRIUS pipeline software produces median combined images that comprise 10 dithered 5\,s exposures. These combined images are ideally free from the spurious noise caused by the presence of bad/hot pixels and/or by cosmic ray events. 

\subsection{Astrometry}
The celestial coordinates of the sources detected in our survey are calculated by referencing their positions to the 2MASS point source catalogue. This process involves the following steps for each field of view:
\begin{enumerate}
\item The equatorial coordinates ($\alpha_i$, $\delta_i$) of 2MASS sources are converted to ($X_i$, $Y_i$) in the World Coordinate System.
\item Bright sources are extracted and their pixel coordinates ($x_j$, $y_j$) measured.
\item A triangle matching technique is used to relate ($X_i$, $Y_i$) and ($x_j$, $y_j$) with the IRAF\footnote{IRAF is distributed by the National Optical Astronomy Observatories, which are operated by the Association of Universities for Research in Astronomy, Inc., under cooperative agreement with the National Science Foundation.} task XYXYMATCH.
\item The transformation matrix to relate ($x_j$, $y_j$) and ($\alpha_i$, $\delta_i$) is calculated using the matched pairs with CCMAP/IRAF. We find more than a hundred matched pairs in all fields of view for all wavebands to calculate the matrix. The individual matrices are used to derive the coordinates of all sources detected in the images.
\end{enumerate}

The positional differences between the 2MASS and the resultant fit coordinates are always quite small. The root mean square values of the positional differences between them are typically 0.02$^{\prime\prime}$ to 0.05$^{\prime\prime}$ for all wavebands. The absolute astrometric precision of the 2MASS point source catalogue is about 0.07$^{\prime\prime}$ (\citealt{skrutskie2006}) and our fit coordinates should have an accuracy of the same order. The equatorial coordinates of the 2MASS point source catalogue are based on the International Celestial Reference System (ICRS). Hence, our fitted coordinates refer to the ICRS.

\subsection{Photometry}
\label{sec:photo}
There are more than a hundred independent images for each field of view (see section \ref{sec:strategy}) taken at various dates and times. Let ${}^{f}N_{\textrm{obs}}$ be the number of the independent images in a field $f$. Poor images have been excised from the following analyses based on the star count in each image. First,  the number of stars detected in each image is counted and the maximum star count, ${}^{f}n_{\textrm{max}}$, over ${}^{f}N_{\textrm{obs}}$ images is found. Then, images with star counts fewer than $0.3 \times {}^{f}n_{\textrm{max}}$ are regarded as poor images. The number of available exposures, ${}^{f}N_{\textrm{exp}}$, differs from one field to another due to temporal changes of observing conditions (weather, seeing, sky, etc.), and ${}^{f}N_{\textrm{exp}} \leq  {}^{f}N_{\textrm{obs}}$ by definition.

As is described in Section \ref{sec:detection}, we use an image subtraction technique to detect light variations. It measures the differential brightness from a certain reference brightness measured in a reference image of a field. Therefore it is technically not necessary to do photometry on all ${}^{f}N_{\textrm{exp}}$ images. Instead, it is sufficient to do photometry on a reference image of each field.  The photometric reference image for each field is made by combining the 10 best-seeing images with typical seeing of 1 arcsec, after eliminating the shift and rotation between images as well as the differences in seeing and backgrounds. In this process the 10 images are filtered using 3-$\sigma$ rejection from the median. We use these combined images for photometric and positional references for each field. In addition to the reference image, we nonetheless perform photometry on all ${}^{f}N_{\textrm{exp}}$ images in each field. This is for evaluating variability (see Section~\ref{sec:evaluation}).

We developed point spread function (PSF) fitting photometry software working under IRAF. This process involves the following steps:
\begin{enumerate}
\item DAOFIND is used to extract point-like sources whose fluxes are more than 3-$\sigma$ above the background noise level.
\item Aperture photometry is performed on the extracted sources using an aperture radius of 7 pixels. The inner radius of the sky annulus is the same as the aperture radius and the width of the sky annulus is 10 pixels. 
\item Several isolated (i.e., no bright sources within 7 pixels) point sources with moderate flux (i.e., unsaturated with good S/N ratio) are selected from the result of step (ii). At least 25 such ``good" stars are selected.
\item Since the shapes of the PSFs can vary from image to image the stars selected in step (iii) are used to construct a model PSF for each image. We let the PSF/DAOPHOT package choose a best fitting function by trying several different types.
\item The PSF fitting photometry is performed on the extracted sources in step (i) using  ALLSTAR. We assume that the PSF is constant over an image.
\end{enumerate}
This photometric process yields arbitrary instrumental magnitudes for each source that have yet to be calibrated.

\subsubsection{Photometric calibration}
We reference the IRSF Magellanic Clouds Point Source Catalog (IRSFMCPSC, \citealt{kato}) to convert the instrumental magnitudes to the calibrated ones. The IRSFMCPSC is calibrated with Las Campanas Observatory (LCO) standards from \citet{persson}. Therefore our photometric zero points are also based on the LCO standards. Because our data are taken with the same instrument (i.e., IRSF/SIRIUS) as was used to collect data to make the IRSFMCPSC, we assume a simple linear equation between the instrumental and calibrated magnitude,
\begin{equation}
{}^{\lambda}m_{\textrm{calibrated}} = {}^{\lambda}m_{\textrm{instrumental}} + {}^{\lambda}\textrm{offset},
\label{conv}
\end{equation}
where $\lambda$ denotes the $J$, $H$ and $K_s$ filters. Conversion to other systems such as the 2MASS system are given elsewhere (e.g., \citealt{kato}). The conversion offset in equation \ref{conv} is determined for each image by taking the difference between the instrumental magnitude and the catalogue magnitude from the IRSFMCPSC. In this process, sources detected well away from the detector edges and correlated catalogue sources with 2MASS quality flag A are used. Then we take a weighted average of the difference and used it as the conversion offset. The weighting factors are the inverse square of the total errors, calculated by combining the catalogue- and instrumental magnitude errors. Outliers are rejected by an iterative 2-$\sigma$ clipping algorithm from the weighted average. Typically more than a hundred sources are used to calculate the final offset value. The standard error of the offset is important in determining the final photometric accuracy in individual fields of view. The standard error of the offset in each field is found to be typically 0.001 to 0.002 mag for all wavebands, which should be considered to be the systematic error of the photometry.
 
We apply the individual offset value and its corresponding standard error for each field of view to calibrate the instrumental magnitudes. Hereafter we call these magnitudes ``calibrated magnitudes".

\subsubsection{Evaluation of photometry using sources in overlapping fields}
We mosaiced the 1$^\circ$$\times$1$^\circ$ area in the central part of the SMC with 81 fields of view. Each field (7.7$^\prime$$\times$7.7$^\prime$) overlaps at its edges with adjacent fields by about 0.7 arcmin. Sources falling in these overlap regions have multiple photometric measurements that should be consistent as long as the sources are not variable. We evaluate our photometry of the reference data by checking the consistency of calibrated magnitudes extracted from each of these measurements on photometric reference images. 

\begin{figure}
\centering
\includegraphics[scale=0.33,bb=0 0 504 720,angle=-90]{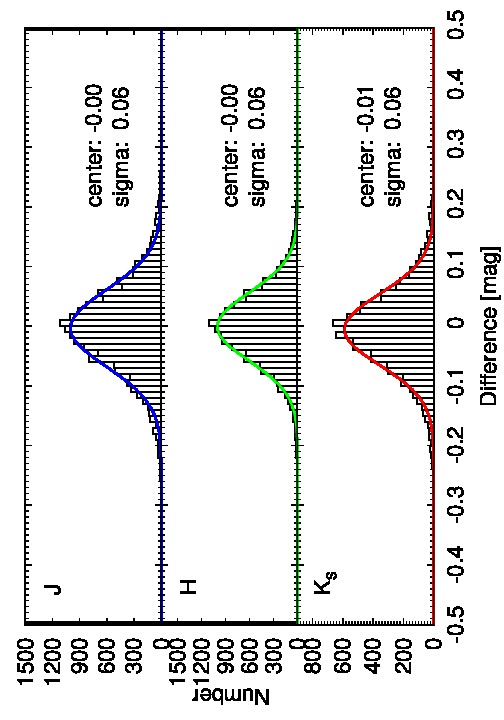}
\caption{The distribution of photometric differences for all sources with S/N$>20$ in overlap regions between adjacent fields. The bin size is 0.01 mag. The thick solid lines show the results of fitting a Gaussian profile to the data.} 
\label{difference}
\end{figure}

The distribution of photometric differences for all high signal-to-noise (S/N $>20$) sources common to adjacent pairs is shown in Fig.~\ref{difference}. We fit a Gaussian profile to estimate the mean difference and its standard deviation. They are, $-0.00\pm0.06$, $-0.00\pm0.06$, and $-0.01\pm0.06$ mag in $J$, $H$ and $K_s$, respectively. These results suggest that our photometric calibration is good and uniform across the survey field. The dispersions seen in the figure arises mainly from photometric random errors and internal dispersions in the IRSFMCPSC, which is about 0.04 mag (\citealt{kato}).

\begin{figure}
\centering
\includegraphics[scale=0.33,bb=0 0 504 720,angle=-90]{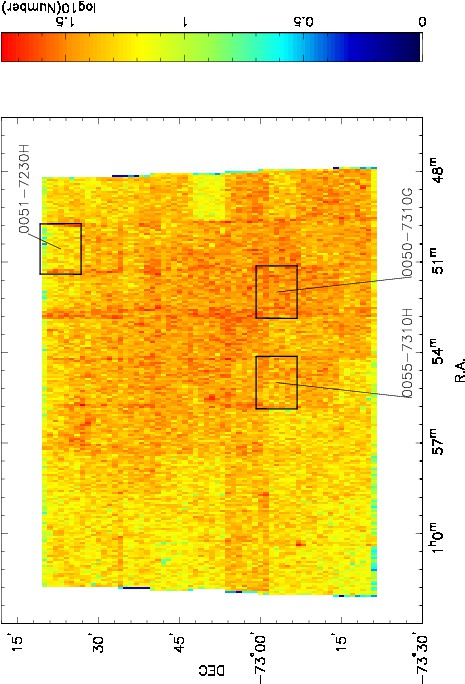}
\caption{Star density map (detected in the $H$ band and binned by 1 arcmin$^2$) of the survey area in the SMC. The positions of the three test fields used for the completeness analyses are indicated.} 
\label{densitymap}
\end{figure}

\begin{figure}
\centering
\includegraphics[scale=0.33,bb=0 0 504 720,angle=-90]{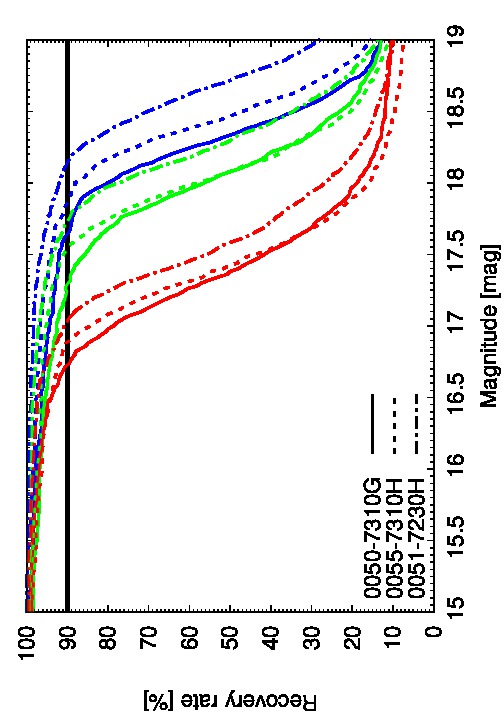}
\caption{Recovery rate vs. magnitude diagram for three fields of view with different stellar number densities. The differences in line types show different fields of view as indicated in the figure. The colours of the lines correspond to different filter bands, blue for $J$, green for $H$, and red for $K_s$, respectively.} 
\label{completeness}
\end{figure}

\begin{table*}
\centering
\caption{Source number densities, 90\% completeness limits, and 10\% detection limits in the three selected fields of view.}
\label{comptable}
\begin{tabular}{lrrcccccccccc}
\hline
\multicolumn{1}{c}{Items} & \multicolumn{9}{c}{Field of view} \\
\cline{2-10}
\multicolumn{1}{c}{ } & \multicolumn{3}{c}{0051--7230H} & \multicolumn{3}{c}{0055--7310H} & \multicolumn{3}{c}{0050--7310G} \\
\multicolumn{1}{c}{ } & \multicolumn{1}{c}{$J$} & \multicolumn{1}{c}{$H$} & \multicolumn{1}{c}{$K_s$} & \multicolumn{1}{c}{$J$} & \multicolumn{1}{c}{$H$} & \multicolumn{1}{c}{$K_s$} & \multicolumn{1}{c}{$J$} & \multicolumn{1}{c}{$H$} & \multicolumn{1}{c}{$K_s$} \\
\hline
density [sources/arcmin$^2$] & 57 & 55 & 50 & 74 & 73 & 61 & 84 & 96 & 69  \\
90\% completeness limit [mag]& 18.14 & 17.72 & 17.02 & 17.84 & 17.52 & 16.88 & 17.64 & 17.28 & 16.72 \\
10\% detection limit [mag]& 18.4 & 17.8 & 16.9 & 18.3 & 17.6 & 16.7 & 18.2 & 17.4 & 16.5 \\
\hline
\end{tabular}
\end{table*}

\subsubsection{Completeness of point source detections}
In order to estimate the detection completeness of our source extraction we use an artificial star technique. We add 900 artificial point sources (i.e., stars) at a given magnitude to the photometric reference images. The extra stars are distributed on a 30$\times$30 grid with a spacing of 30 pixels. The sources are extracted from the new artificial image in the same way as is described in Section~\ref{sec:photo}. The list of input artificial stars is then cross-identified with the list of detected stars to examine how many artificial stars are successfully extracted. These processes are repeated by varying the input source magnitude in steps of 0.02 mag. We define the ``90 percent completeness limit'' as the magnitude at which 90 percent of the added artificial stars are recovered.

To estimate the effect of source number density on the completeness analyses, three 7.7$^{\prime}$$\times$7.7$^{\prime}$ areas in the SMC are selected as test fields. Fig.~\ref{densitymap} is the star density map (number of stars detected in the $H$ band in 1 arcmin$^2$ bins) of the observed 1$^\circ$$\times$1$^\circ$ area with the three test fields indicated. One is the most crowded field (0050--7310G, $\sim$ 96 stars/arcmin$^2$ in $H$), another is moderately populated (0055--7310H, $\sim$ 70 stars/arcmin$^2$ in $H$) and the last one is sparsely populated (0051--7230H, $\sim$ 55 stars/arcmin$^2$ in $H$). The results of the analyses are summarized in Table~\ref{comptable} and in Fig.~\ref{completeness}. These indicate that the 90 percent completeness limits are deeper in the underpopulated fields of view and that they can differ up to about 0.50, 0.44, and 0.30 mag for $J$, $H$ and $K_s$, respectively, between the sparsely and densely crowded extremes.

\subsubsection{Detection limits}
\begin{figure}
\centering
\includegraphics[scale=0.33,bb=0 0 504 720,angle=-90]{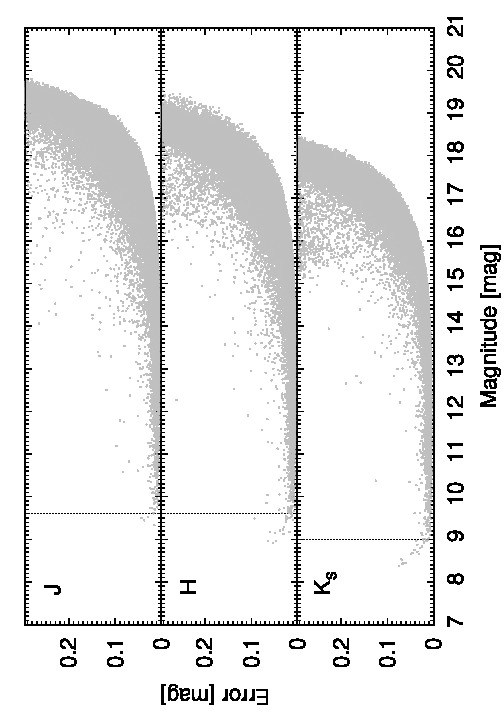}
\caption{The distributions of photometric error as a function of calibrated magnitudes for all sources detected in the surveyed area. The vertical thin dotted lines indicate the saturation limits.} 
\label{errdist}
\end{figure}

The distributions of photometric error\footnote{The photometric error includes errors in the magnitude calculated by IRAF.} versus calibrated magnitude for the all sources detected in the photometric reference images are shown in Fig.~\ref{errdist}. It is clear that the brightest sources are affected by deviations from the linear response of the SIRIUS detectors. The deviation becomes larger than 2 per cent for sources brighter than $J=9.6$, $H=9.6$, and $K_s=9.0$ mag, respectively (hereafter referred as ``saturation limits''). Certainly, these values depend on the observational conditions, such as seeing size and sky background brightness. Any sources brighter than these limits are not included in the catalogues published with this paper, and are excluded from the following discussions. Fig.~\ref{errdist} also illustrates the typical photometric errors of our survey as a function of source luminosity.

We also define the 10\% detection limit as the magnitude at which the photometric errors of the sources exceeds 0.109 mag. The 10\% detection limits are calculated for the aforementioned three different population density fields, and are summarized in the Table~\ref{comptable}. The results indicate that the 10\% detection limits depend on the source density.

\subsection{Detection of variable source candidates}
\label{sec:detection}
\subsubsection{Image subtraction}
We use the image subtraction package, ISIS.V2.2, to detect variable source candidates. The image subtraction method can find variable sources even in very crowded fields and its efficiency in detecting variables is evidenced by previous and ongoing surveys (e.g., OGLE and MOA). Details of the image subtraction technique are given in \citet{alard1998} and \citet{alard2000} and only an outline of the technique is described here. 

First, images are shifted and rotated to match the photometric reference image, which is also used as the positional reference. Basically, the image subtraction technique stands on the idea that two images (an image and a reference image) taken under different conditions (i.e., seeing and sky background) are related by the following equation:
\begin{equation}
I_i(x,y) = f_i(R(x,y)) + \textrm{bg}_i(x,y),
\end{equation}
which transforms the PSF with smaller full-width at half maximum of the reference image $R$ to the PSF of an image $I_i$, where $i$ denotes the $i$-th observation, $f_i$ is the convolution function for $i$-th observation and bg$_i$$(x,y)$ is the differential sky background between the two images. We assume that there are $N_{\textrm{exp}}$ available exposures in all ($i = 1, N_{\textrm{exp}}$). After the convolution the differential image, $D_i(x,y)$, can be computed as:
\begin{equation}
D_i(x,y) = I_i(x,y) - [ f_i(R(x,y)) + \textrm{bg}_i(x,y) ].
\end{equation}
Then in the ideal case all the pixels except those of variable sources should have values of zero in the differential image, $D_i(x,y)$. Finally a variance image is created by calculating the variance of each pixel over $i$ in the differential images: 
\begin{equation}
V(x,y) = \frac{\sum_{i=1}^{N_{\textrm{exp}}} (D_i(x,y) - \overline{D_i(x,y)})^2 }{N_{\textrm{exp}}},
\end{equation}
where
\begin{equation}
\overline{D_i(x,y)} = \frac{\sum_{i=1}^{N_{\textrm{exp}}} D_i(x,y)}{N_{\textrm{exp}}}.
\end{equation}
If a variable source is found at position $(x, y)$ then the variance of the pixel should be large. In this way variable sources stand out and non-variables are invisible on the variance image. We consider candidate variable sources to be those stars that have signals 3-$\sigma$ above the local background in the variance image. The positions of the variable source candidates are measured simultaneously. The number of variable source candidates is summarized in Table~\ref{candidate}.

\subsubsection{Converting ADU to magnitude}
The differential image technique calculates flux differences in Astronomical Data Units (ADU). Therefore it is necessary to convert ADU to calibrated magnitudes for practical use. After we obtain the instrumental magnitudes and their corresponding ADU fluxes in the photometric reference image we can convert ADU light curves into calibrated magnitudes through the steps described below. First we define the following values:
\begin{itemize}
\item ${}^{\lambda}I^{j}_0$: the flux at a given waveband $\lambda$ of the $j$-th variable source candidate in the reference image in ADU.
\item ${}^{\lambda}I^{j}_i$: the flux at a given waveband $\lambda$ of the $j$-th variable source candidate in the $i$-th image in ADU.
\end{itemize}
The ISIS software gives differential fluxes of variable sources relative to their fluxes in the reference image, {\it ie}, $\Delta{}^{\lambda}I^{j}_i = {}^{\lambda}I^{j}_0 - {}^{\lambda}I^{j}_i$. To convert these values to magnitudes we need both the instrumental magnitude, ${}^{\lambda}m^{j}_0$, and the flux, ${}^{\lambda}I^{j}_0$, of each variable source candidate in the photometric reference image \citep{benko2001}. Candidate variables in the photometric reference image are selected by the nearest neighbour search method. A search radius of 4 pixel is used. If several sources are present within the search radius on the reference image, we choose the closest one and set the proximity flag (see Section~\ref{sec:timeseries}). The search radius (corresponding to about $1.8^{\prime\prime}$) was chosen by experiment, and takes into account variations in observing conditions that can make accurate position determination in the variance image difficult. Then we calculate the calibrated magnitudes, ${}^{\lambda}m^{j}_i$, of the $j$-th variable source candidate in the $i$-th image through the ordinal formula,
\begin{equation}
\label{henkan}
{}^{\lambda}m^{j}_i = -2.5 \log_{10} \left( \frac{{}^{\lambda}I^{j}_0 - \Delta {}^{\lambda}I^{j}_i}{{}^{\lambda}I^{j}_0} \right) + {}^{\lambda}m^{j}_0 + \textrm{conversion offset},
\end{equation}
where the conversion offset is derived as in Section~\ref{sec:photo} for each waveband and field of view.

\subsection{Evaluating variability}
\label{sec:evaluation}
\subsubsection{Variability indices and false alarm probability}
A key characteristic of our data is that they are taken simultaneously in the $J$, $H$ and $K_s$ bands. Such a multi-band simultaneous photometric data set is ideal for searching for variability through correlations between photometric fluctuations at different wavelengths (\citealt{welch1993}; \citealt{stetson1996}). This robust technique for evaluating stellar variability is further improved by \cite{lopes2015} and \cite{lopes2016} to handle panchromatic flux correlations. They defined several statistical variability indices, but the main ones used in this work are the so-called Stetson index ``$J_{\textrm{WS}}$'' and the Ferreira Lopes indices, ``$I^{(s)^\prime} _{\textrm{pfc}}$'' and ``$I^{(s)} _{\textrm{fi}}$''. 

The Stetson index, $J_{\textrm{WS}}$, is calculated as
\begin{equation}
{}^{\lambda_1,\lambda_2}J_{\textrm{WS}} = \sum_{i=1}^{N_\textrm{pair}} {}^{\lambda_1,\lambda_2}\Gamma_{i} \sqrt{|{}^{\lambda_1}\delta_{i} {}^{\lambda_2}\delta_{i}|},
\end{equation}
where $N_\textrm{pair}$ is the number of simultaneous observation pairs in two different wavebands, $\lambda_1$ and $\lambda_2$, the ${}^{\lambda}\delta_{i}$ is the normalized residual for a given waveband, $\lambda$, computed as
\begin{equation}
{}^{\lambda}\delta_{i} = \sqrt{\frac{N_\textrm{pair}}{N_\textrm{pair}-1}} \left( \frac{{}^{\lambda}m_{i} - {}^{\lambda}\mu}{{}^{\lambda}e_{i}} \right),
\end{equation}
where ${}^{\lambda}m_{i}$ and ${}^{\lambda}e_{i}$ are the time-series photometric measurements and their corresponding errors for a given waveband, respectively, and ${}^{\lambda}\mu$ is the weighted mean, ${}^{\lambda}m_{i}$. At first, a simple mean of ${}^{\lambda}m_{i}$ is used for ${}^{\lambda}\mu$. Then a weighting factor, ${}^{\lambda}g_{i}$, which is defined in \cite{stetson1996} as
\begin{equation}
{}^{\lambda}g_{i} = \left[ 1 + \left( \frac{|{}^{\lambda}\delta_i|}{2} \right)^2 \right]^{-1},
\end{equation}
is calculated and ${}^{\lambda}\mu$ is redetermined with those weights. The procedure is iterated until ${}^{\lambda}\mu$ stabilizes. This helps to reduce the influence of any outliers in a set of measurements. Also, the correction factor, ${}^{\lambda_1,\lambda_2}\Gamma_{i}$, is given by
\[
  {}^{\lambda_1,\lambda_2}\Gamma_{i} = \begin{cases}
    1 & (\textrm{if} ~ {}^{\lambda_1}\delta_{i} \cdot {}^{\lambda_2}\delta_{i} > 0) \\
    -1 & (\textrm{otherwise.})
  \end{cases}
\]

The Ferreira Lopes index, $I^{(s)^\prime} _{\textrm{pfc}}$ (in our case, the combination type $s$ is equal to 3), is calculated as
\begin{equation}
I^{(3)^\prime} _{\textrm{pfc}} = \frac{1}{N_\textrm{pair}} \sum_{i=1}^{N_\textrm{pair}}  {}^{\lambda_1,\lambda_2,\lambda_3}\Lambda_{i} \sqrt[3]{|{}^{\lambda_1}\delta_{i} {}^{\lambda_2}\delta_{i} {}^{\lambda_3}\delta_{i}|},
\end{equation}
with ${}^{\lambda_1,\lambda_2,\lambda_3}\Lambda_{i}$ defined as
\[
  {}^{\lambda_1,\lambda_2,\lambda_3}\Lambda_{i} = \begin{cases}
    1 & (\textrm{if} ~ {}^{\lambda_1}\delta_{i} > 0, {}^{\lambda_2}\delta_{i} > 0, {}^{\lambda_3}\delta_{i} > 0) \\
    1 & (\textrm{if} ~ {}^{\lambda_1}\delta_{i} < 0, {}^{\lambda_2}\delta_{i} < 0, {}^{\lambda_3}\delta_{i} < 0) \\
    -1 & (\textrm{otherwise.})
  \end{cases}
\]
The other Ferreira Lopes index, $I^{(s)} _{\textrm{fi}}$, is calculated as
\begin{equation}
I^{(3)} _{\textrm{fi}} = 0.5 \left[ 1 + \frac{1}{N_\textrm{pair}} \sum_{i=1}^{N_\textrm{pair}}  {}^{\lambda_1,\lambda_2,\lambda_3}\Lambda_{i}
 \right],
\end{equation}
which takes values from 0 to 1. Refer to the papers mentioned above for the original definitions of the variability indices. 

For purely random photometric errors, photometric fluctuations at different wavelengths should be uncorrelated, and their ${}^{\lambda_1,\lambda_2}J_{\textrm{WS}}$ or $I^{(3)^\prime} _{\textrm{pfc}}$ should tend to zero in the limit of a large number of observations. The $I^{(3)} _{\textrm{fi}}$ index provides a measure of signal correlation strength, with 1 being the perfect correlation over all wavebands observed. For our data, the number of pairs, $N_\textrm{pair}$, is typically more than one hundred. A minimum condition of $N_\textrm{pair} > 10$ is set for calculating these variability indices. Note that the number of pairs, $N_\textrm{pair}$, is equal to or less than the number of available exposures, $N_{\textrm{exp}}$, because only data brighter than 19.0, 18.5, and 17.5 mag in the $J$, $H$ and $K_s$ bands, respectively, are used. These limits roughly correspond to the brightnesses where their typical photometric errors exceed 0.3 mag (see Fig.~\ref{errdist}). Hereafter these limits are referred to as ``faint limits''. Also, to minimize adverse effects from possible spikes in the data, up to five extreme outlier data points are identified by a 2 sigma-clipping algorithm. These possible spikes as well as data points fainter than the faint limit are ignored in the process of calculating the above indices. 

During the calculation of the variability indices, we also computed the false alarm probability, FAP (how often a certain signal is observed just by chance), by using so-called Monte Carlo or bootstrap simulations. At first the time-series data are randomly shuffled keeping the times of observations fixed. Then the variability index of the randomly shuffled data is calculated in the same manner as before. The data are randomly reshuffled and the process is repeated. This procedure is repeated 10,000 times ($= N_{\textrm{total}}$) and the number of times the absolute value of the calculated index exceeds the original one ($= N_{\textrm{chance}}$) is recorded. Then we calculate the FAP as $\textrm{FAP} = N_{\textrm{chance}}/N_{\textrm{total}}$. 

\begin{figure}
\centering
\includegraphics[scale=0.33,bb=0 0 504 720,angle=-90]{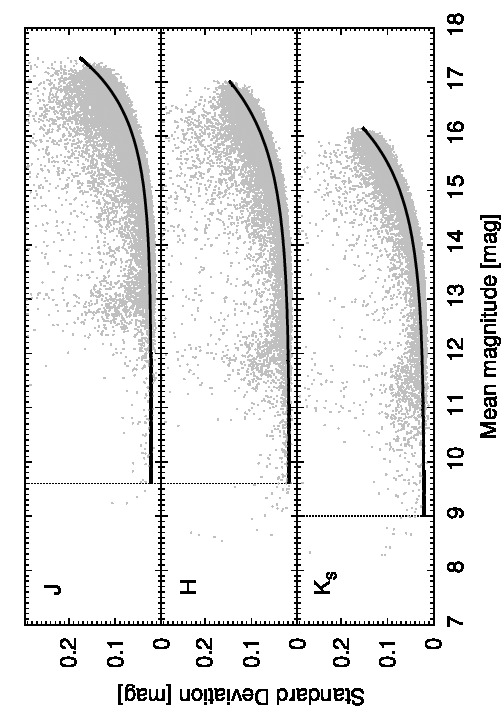}
\caption{The standard deviation of ${}^{\lambda}N_\textrm{det}$ magnitudes is plotted against the mean magnitudes for all stars detected regardless of variability. Only stars with ${}^{\lambda}N_\textrm{det} > 0.5{}^{\lambda}N_\textrm{exp}$ are shown. The solid lines connect the median standard deviations at a given mean magnitude calculated for every 0.1 mag interval with a width of $\pm$0.2 mag. The dashed lines are the fits to the points of a function of the form ${}^{\lambda}\textrm{SD}(X) = {}^{\lambda}a + {}^{\lambda}b \exp\left( {}^{\lambda}c {}^{\lambda}X\right)$, where ${}^{\lambda}X$ is the mean magnitude; these define the limits separating variable from non-variable sources. See text for details.} 
\label{sdplot}
\end{figure}

\begin{table}
\centering
\caption{The best fit parameters.}
\label{sdfit}
\begin{tabular}{ccccc}
\hline
\multicolumn{1}{c}{Waveband} & \multicolumn{1}{c}{${}^{\lambda}a$} & \multicolumn{1}{c}{${}^{\lambda}b$} & \multicolumn{1}{c}{${}^{\lambda}c$} & \multicolumn{1}{c}{${}^{\lambda}\chi^2$}\\
\hline
$J$   & 0.022 & 1.174$\times$10$^{-8}$ & 0.940 & 0.003 \\
$H$   & 0.017 & 2.818$\times$10$^{-7}$ & 0.766 & 0.001 \\
$K_s$ & 0.020 & 3.230$\times$10$^{-7}$ & 0.800 & 0.001 \\
\hline
\end{tabular}
\end{table}

\subsubsection{Standard deviation of observed magnitudes}
Standard deviations of observed magnitudes are also calculated for each candidate. Again, to minimize adverse effects from possible spikes in the data, up to five extreme outlier data points are identified by a 2-sigma clipping algorithm and are ignored. Here we define ${}^{\lambda}N_\textrm{det}$ to denote the number of detections that passed the screening and are also brighter than the faint limit over ${}^{\lambda}N_\textrm{exp}$ available exposures in a given waveband, $\lambda$. ${}^{\lambda}N_\textrm{det}$ can be equal to or less than ${}^{\lambda}N_\textrm{exp}$, and can be different for each source. Note again that the possible spikes and data points fainter than the faint limit are rejected only for the purpose of calculating the standard deviation, and are present in the published time-series data. Then we calculated the standard deviations of ${}^{\lambda}N_\textrm{det}$ magnitudes for sources with ${}^{\lambda}N_\textrm{det} > 0.5{}^{\lambda}N_\textrm{exp}$ regardless of variability. The calculated standard deviations are plotted against the corresponding mean magnitudes in Fig.~\ref{sdplot}. Not surprisingly, the standard deviation tends to become large as luminosity decreases. Obviously, large amplitude variables such as Mira-like variables blur the trend, but the majority of the sources plotted in the figure are not variable and we assume that the standard deviations arise mainly from photometric errors. 

To formulate the trend we first calculated the median of the standard deviations, ${}^{\lambda}\textrm{SD}({}^{\lambda}X)$, for every 0.1 mag interval with a width of $\pm$0.2 mag. Then, an exponential curve of the form:
\begin{equation}
\label{variablecondition}
{}^{\lambda}\textrm{SD}({}^{\lambda}X) = {}^{\lambda}a + {}^{\lambda}b \exp\left( {}^{\lambda}c {}^{\lambda}X\right)
\end{equation}
was fitted by the least squares method, where ${}^{\lambda}a$, ${}^{\lambda}b$, and ${}^{\lambda}c$ are free parameters, and ${}^{\lambda}X$ is the mean magnitude at which the corresponding ${}^{\lambda}\textrm{SD}({}^{\lambda}X)$ was calculated. We imposed a constraint that the standard deviation be a monotonically increasing function of mean magnitude, {\it ie}, ${}^{\lambda}\textrm{SD}({}^{\lambda}X_1) < {}^{\lambda}\textrm{SD}({}^{\lambda}X_2)$ if ${}^{\lambda}X_1 < {}^{\lambda}X_2$ and ${}^{\lambda}\textrm{SD}({}^{\lambda}X_1) = {}^{\lambda}\textrm{SD}({}^{\lambda}X_2)$ otherwise. The resultant fitted curves are shown by thick lines in Fig.~\ref{sdplot}. The best-fitting parameter values are tabulated in Table~\ref{sdfit} together with the corresponding $\chi^2$ values. The derived formulae define thresholds to separate real variables from non-variable sources, and are also used for identifying variable sources.

\subsubsection{Identifying variable sources}
Following the procedure described in section \ref{sec:detection}, we found more than 6\,0000 candidates in the survey field. We take the following steps to reject false variability candidates. In the following, let ${}^{\lambda}\mu^j$ and ${}^{\lambda}\textrm{sd}^j$ denote the mean and the standard deviation, respectively, of the ${}^{\lambda}m_{i}^j$ in Equation~\ref{henkan} calculated with data brighter than the faint limit and that passed the up-to-five extreme outliers rejection procedure described above.

\paragraph{Rejecting foreground stars with relatively high proper motion}
\label{sec:hpm}
We have been monitoring the survey area for more than a decade. Therefore, for some foreground stars in the survey area, the apparent positional changes over this period can become large enough to detect. Such a relatively high proper motion (HPM) star can be misclassified as a variable star in our method of finding variables and should be rejected. We cross-matched the variable source candidates with the second data release (DR2) of $Gaia$ catalogue (GAIA DR2, \citealt{gaia2018}; \citealt{lindegren2018}) with a tolerance radius of 1.0 arcsec and reject sources with total proper motion ($\sqrt{ \mu_\alpha^2 \cos^2 \delta + \mu_\delta^2}$) larger than 22.5 mas/year. The threshold value of 22.5 mas/year was chosen to reject stars that would have moved more than half a pixel\footnote{Recall that the SIRIUS camera has a pixel field of view of about 0.45 arcsec.} in 10 years. We further cross-matched the variable source candidates with the OGLE catalog of high proper motion stars towards the Magellanic Clouds (\citealt{soszynski2002} and \citealt{poleski2011}) with a tolerance radius of 1.0 arcsec and removed ones with cross-identifications. We checked the light curves of all matched HPM star candidates. Most of them have distinctive light curves of false variables (e.g., linearly increasing or decreasing light curve), and none of the rest have meaningful light curves. In these screening processes 698 sources were rejected. Further investigation of the individual light curves revealed another 47 stars with linearly changing light curves. These all have $Gaia$ proper motions $>$4 mas/yr and parallaxes $>$0.3 mas, and have been removed from our catalogue of variables.

\paragraph{Candidates with 3-band detections}

\begin{figure}
\centering
\includegraphics[scale=0.33,bb=0 0 504 720,angle=-90]{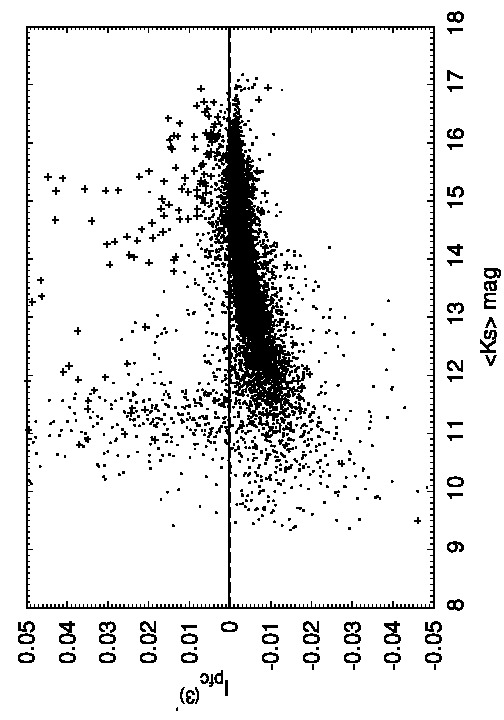}
\includegraphics[scale=0.33,bb=0 0 504 720,angle=-90]{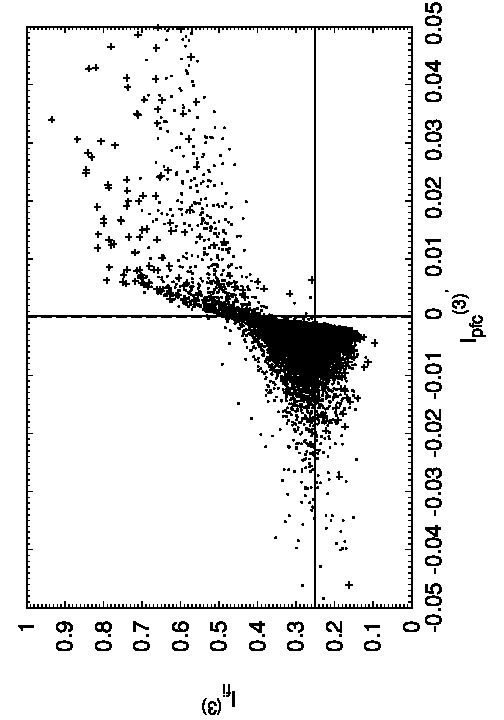}
\caption{Upper panel: Ferreira Lopes variability index, $I^{(3)^\prime} _{\textrm{pfc}}$, as a function of weighted magnitude mean  ${}^{K_s}\mu$ for variable source candidates. Lower panel: $I^{(3)} _{\textrm{fi}}$ index against $I^{(3)^\prime} _{\textrm{pfc}}$ index for variable source candidates. Small dots and crosses represent false alarm probability greater than or equal to 1/10000 and less than 1/10000, respectively. The dashed line at $I^{(3)^\prime} _{\textrm{pfc}} = 0$ shows the expected value of the index for random noise. The solid line represents the cutoff value adopted in identifying variable sources. For clarity, 152 sources with $\left| I^{(3)^\prime} _{\textrm{pfc}} \right| \geq 0.05$ are not shown.}
\label{ferreiralopes}
\end{figure}

For variable candidates with 3-band detections, we use the $I^{(3)^\prime} _{\textrm{pfc}}$, $I^{(3)} _{\textrm{fi}}$, and ${}^{\lambda}\textrm{sd}$ indices for evaluating variability. A variable source candidate is recognized as a real variable if all of the following conditions are satisfied:
\begin{itemize}
\item $I^{(3)^\prime} _{\textrm{pfc}} \geq 0.0002$
\item $I^{(3)} _{\textrm{fi}} \geq 0.25$
\item ${}^{J}\textrm{sd} \geq 2.0\times{}^{J}\textrm{SD}({}^{J}\mu)$ OR ${}^{H}\textrm{sd} \geq 2.0\times{}^{H}\textrm{SD}({}^{H}\mu)$ OR ${}^{K_s}\textrm{sd} \geq 2.0\times{}^{K_s}\textrm{SD}({}^{K_s}\mu)$
\end{itemize}
The upper and lower panels of Fig.~\ref{ferreiralopes} show the Ferreira Lopes variability index, $I^{(s)^\prime} _{\textrm{pfc}}$, as a function of the weighted magnitude mean, ${}^{K_s}\mu$, and the  $I^{(3)} _{\textrm{fi}}$ index against $I^{(3)^\prime} _{\textrm{pfc}}$, respectively. Cutoff values, above which a variable source candidate is considered to be real were estimated by visually examining the light curves as a function of indices while taking into account the FAP. By definition, candidates with large negative $I^{(s)^\prime} _{\textrm{pfc}}$ values could also be real variables. However, experiments showed that they are likely to be false alarms, probably resulting from systematic photometric errors in a given waveband while photometry of the other two wavebands correlates. Therefore we restrict ourselves to stars with positive index values. The resultant cutoff values are indicated by solid lines in the figures. 

\paragraph{Candidates with 2-band detections}

\begin{figure}
\centering
\includegraphics[scale=0.32,bb=0 0 504 720,angle=-90]{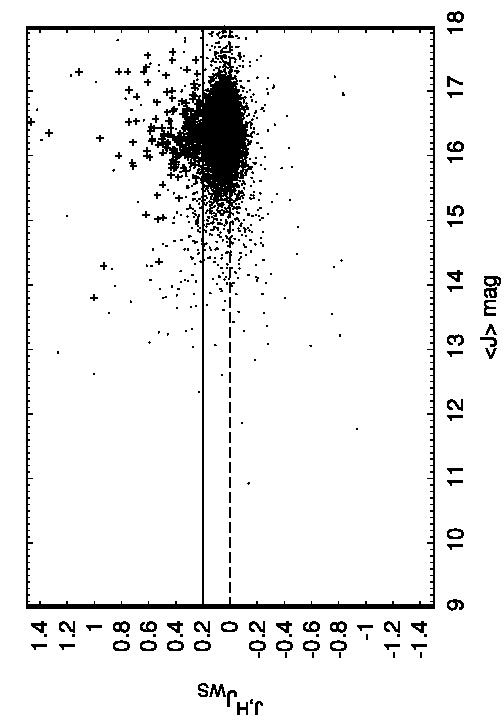}
\includegraphics[scale=0.32,bb=0 0 504 720,angle=-90]{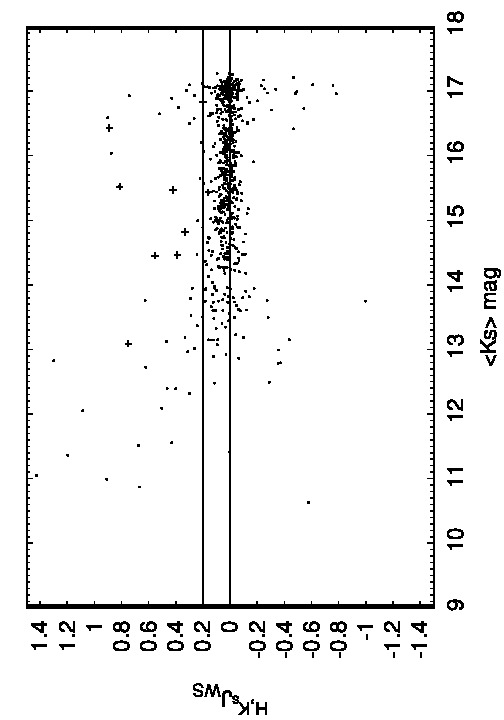}
\includegraphics[scale=0.32,bb=0 0 504 720,angle=-90]{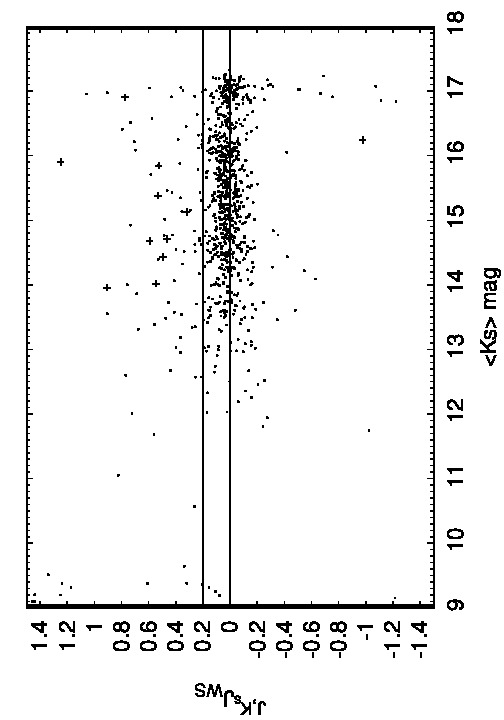}
\caption{Stetson variability index, ${}^{\lambda_1,\lambda_2}J_{\textrm{WS}}$, for candidates with 2-band detections is plotted as a function of weighted magnitude mean ${}^{J}\mu$ or ${}^{K_s}\mu$. Symbols and lines are as in Fig.~\ref{ferreiralopes}. For clarity, 34, 12, and 21 stars with $\left| {}^{\lambda_1,\lambda_2}J_{\textrm{WS}} \right| \geq 1.5$ are not shown for $J\&H$, $H\&K_s$, and $J\&K_s$ pairs, respectively.}
\label{stetson}
\end{figure}

For variable candidates with 2-band detections, we use the ${}^{\lambda_1,\lambda_2}J_{\textrm{WS}}$ and ${}^{\lambda}\textrm{sd}$ indices for evaluating variability. A variable source candidate is recognized as a real variable if both of the following conditions are satisfied:
\begin{itemize}
\item ${}^{\lambda_1,\lambda_2}J_{\textrm{WS}} \geq 0.2$
\item ${}^{\lambda_1}\textrm{sd} \geq 2.0\times{}^{\lambda_1}\textrm{SD}({}^{\lambda_1}\mu)$ AND ${}^{\lambda_2}\textrm{sd} \geq 2.0\times{}^{\lambda_2}\textrm{SD}({}^{\lambda_2}\mu)$ 
\end{itemize}
Fig.~\ref{stetson} shows the distribution of the ${}^{\lambda_1,\lambda_2}J_{\textrm{WS}}$ index. The meaning of the marks and lines is the same as in Fig.~\ref{ferreiralopes}. As is the case with the $I^{(3)^\prime} _{\textrm{pfc}}$ index, candidates with large negative ${}^{\lambda_1,\lambda_2}J_{\textrm{WS}}$ values could also be real variables. Again, however, experiments showed that they are likely to be false alarms. Therefore we restrict ourselves to deal with stars with positive index values. Cutoff values were estimated in the same way as for candidates with 3-band detections. 

\paragraph{Candidates with 1-band detections}

For candidates with 1-band detection, first we selected those with ${}^{\lambda}\textrm{sd} \geq 3.0 \times {}^{\lambda}\textrm{SD}({}^{\lambda}\mu)$ and checked all their light curves by eye.

\subsubsection{Summary and comments on variable source identification}
Although more than 99\% of the selected variable sources were detected in more than one waveband, the rest were detected in only one of the three wavebands. Source spectrum, saturation, detector glitches, and so on, are the likely reasons for that. Table~\ref{candidate} shows a breakdown list of the number of variable sources selected by the screening processes described above. Hereafter we refer to these selected variables as ``variable sources''. It should be noted that these sources might include unresolved variable galaxies and QSOs.

Despite our efforts to reject non-physical variable sources, some might have slipped through the screening processes. The main remaining concern is the foreground high proper motion stars without data in the GAIA DR2 catalogue. In our experiments, they have similar light curves, characterised by a steady decline in brightness throughout the period of the survey. In a paper dealing with difference image photometry in the context of gravitational microlensing, \citet{albrow2009} perform a series of experiments that show the effect of measuring a star in the difference image at the wrong coordinates. The effect can be a decrease of about 20\% in apparent brightness for a shift of about 20\% of the FWHM of the images. The declining light curves might then be explained as due to a misplacing of the photometric aperture when measuring the difference images. Any variable sources with steadily declining light curves should be handled with care.

We also notice that there is often a large spread in magnitude in all filters in the data from the later seasons, especially red after the 2013 season. Again, the work by \citet{albrow2009} may be relevant here, in that they showed the error for an off-centre measurement depends on the ratio of the distance off-centre to the FWHM. Also, a deterioration in sensitivity of the SIRIUS camera has been reported (Nakajima, private communication), such that the detection limits in all filters became shallower by about 1 mag over the years since the first season in 2000. The cause of the deterioration is not known. This may also contribute to the spread.

\begin{table}
\caption[]{Number of variable source candidates and sources selected as real. See text for details.}
\label{candidate}
\begin{center}
\begin{tabular}{cccccrr}
\hline
\multicolumn{5}{c}{Wavebands} & \multicolumn{2}{c}{Number} \\
\multicolumn{5}{c}{detected} & \multicolumn{1}{r}{Candidates} & \multicolumn{1}{r}{Selected as real} \\
\hline
$J$ &  & $H$ &   & $K_s$& 24\,561 & 939 \\
$J$ &  & $H$ &   &      & 13\,248 &  90 \\
    &  & $H$ &   & $K_s$&     697 &  17 \\
$J$ &  &     &   & $K_s$&     857 &  13 \\
$J$ &  &     &   &      &  9\,291 &   3 \\
    &  & $H$ &   &      &  8\,692 &   0 \\
    &  &     &   & $K_s$&  3\,615 &   1 \\
\hline
\multicolumn{5}{l}{Total} & \multicolumn{1}{r}{60\,961} & \multicolumn{1}{r}{1\,063} \\
\hline
\end{tabular}
\end{center}
\end{table}

\section{Catalogue Descriptions}
Along with this paper we publish a photometric point source catalogue for the 1$^\circ$$\times$1$^\circ$ survey area and also a variable source catalogue with time-series data. This section describes the features of the catalogues and how they are constructed.

\subsection{Photometric point source catalogue}
When we constructed photometric reference images for each field of view we combined the 10 best-seeing (typically 1 arcsec) images. The 10 images were chosen randomly in time so the photometric results from their combined image will be time-averaged (more specifically, median filtered) over the 10 epochs; these may be useful for certain types of research. In addition to this feature, each of the 10 best-seeing images comprises ten 5\,s exposure images where bright sources will not be saturated. This is an advantage over the \citet{kato} catalogue where the photometry is saturated for sources brighter than about 11 mag. Note also that our photometry  has a S/N comparable to or a bit better than theirs, because of the longer total exposure times (300\,s for the \citet{kato} catalogue and 500\,s for ours). For these reasons we publish the photometric point source catalogue along with the variable source catalogue.

First we simply compiled the photometric results for each field of view. In total 348\,129, 321\,440, and 279\,900 sources are detected in $J$, $H$ and $K_s$, respectively. In this process, sources within a radius of $10^{\prime\prime}$ from the four very bright stars, namely HD5302, CM Tuc, HD5688, and HD6172 were manually deleted because they were heavily affected by the bright haloes of these stars. These simple source lists for each waveband are contaminated by multiply-detected sources that fall in overlapping areas between adjacent fields. We removed such sources based on their spatial proximity ($\left| \Delta r \right| < 1.0^{\prime\prime}$). We adopted the result with the better S/N and discarded the others. This procedure leaves 314\,689, 289\,264, and 252\,140 sources in $J$, $H$ and $K_s$, respectively. Further elimination of saturated sources leaves 314\,685, 289\,239, and 252\,118 sources in $J$, $H$ and $K_s$, respectively. Then the $J$ and $H$ duplication- and saturation-free point source lists were merged using a positional tolerance of $\left| \Delta r \right| < 1.0^{\prime\prime}$. For the matched sources, coordinates were recalculated by taking an average of the coordinates from each band. In the rare cases when more than one source is present within the tolerance radius the closest one was always adopted and the others were listed as solitary sources. The $J$ and $H$ band merged list was further merged with the $K_s$ duplication- and saturation-free point source list in the same way to make a final $J$, $H$ and $K_s$ band merged photometric point source catalogue. Note that the foreground stars rejected from the variable star catalogue (see section~\ref{sec:hpm}) are present in the point source catalogue.

\begin{table*}
\centering
\caption{The first five records that contain meaningful data for all the three wavebands extracted from the photometric point source catalogue. The coordinates are followed by, for each waveband, successive columns listing the calibrated magnitude ${}^{\lambda}m$, the error in the calibrated magnitude ${}^{\lambda}e$, the error in the calibration offset ${}^{\lambda}ec$, and flags for multiple detection ${}^{\lambda}f_m$ and proximity ${}^{\lambda}f_p$. The catalogue is in increasing order of Right Ascension. The full version of the catalogue is available on the MNRAS server.}
\label{photcatalogue}
\begin{tabular}{ccccccccccccccccc}
\hline
\multicolumn{1}{c}{R.A.} & \multicolumn{1}{c}{Declination} & \multicolumn{5}{c}{$J$} & \multicolumn{5}{c}{$H$} & \multicolumn{5}{c}{$K_s$} \\
\multicolumn{2}{c}{} & \multicolumn{1}{c}{$m$} & \multicolumn{1}{c}{$e$} & \multicolumn{1}{c}{$ec$} & \multicolumn{1}{c}{$f_m$} & \multicolumn{1}{c}{$f_p$} & \multicolumn{1}{c}{$m$} & \multicolumn{1}{c}{$e$} & \multicolumn{1}{c}{$ec$} & \multicolumn{1}{c}{$f_m$} & \multicolumn{1}{c}{$f_p$} & \multicolumn{1}{c}{$m$} & \multicolumn{1}{c}{$e$} & \multicolumn{1}{c}{$ec$} & \multicolumn{1}{c}{$f_m$} & \multicolumn{1}{c}{$f_p$} \\
\multicolumn{2}{c}{[degree]} & \multicolumn{3}{c}{[mag]} & \multicolumn{2}{c}{flag} & \multicolumn{3}{c}{[mag]} & \multicolumn{2}{c}{flag} & \multicolumn{3}{c}{[mag]} & \multicolumn{2}{c}{flag} \\
\hline
 11.985383 & -73.303588 & 17.482 &  0.071 &  0.001 & 0 & 1 & 17.009 &  0.065 &  0.001 & 0 & 1 & 17.226 &  0.148 &  0.002 & 0 & 1 \\
 11.986007 & -73.290227 & 17.864 &  0.085 &  0.001 & 0 & 1 & 17.892 &  0.152 &  0.001 & 0 & 1 & 17.676 &  0.222 &  0.002 & 0 & 1 \\
 11.986158 & -73.329104 & 18.204 &  0.099 &  0.001 & 0 & 1 & 17.692 &  0.112 &  0.001 & 0 & 0 & 17.254 &  0.210 &  0.002 & 0 & 0 \\
 11.986227 & -73.256153 & 17.066 &  0.050 &  0.001 & 0 & 0 & 16.597 &  0.048 &  0.001 & 0 & 0 & 16.705 &  0.094 &  0.002 & 0 & 0 \\
 11.986502 & -73.278119 & 17.824 &  0.079 &  0.001 & 0 & 0 & 17.320 &  0.090 &  0.001 & 0 & 0 & 17.127 &  0.136 &  0.002 & 0 & 0 \\
\hline
\end{tabular}
\end{table*}

\subsubsection{The columns of the point source catalogue}
Table~\ref{photcatalogue} shows the first five records that contain meaningful data for all the three wavebands extracted from the photometric point source catalogue as an example. The catalogue is in increasing order of Right Ascension. The full version of the catalogue is available on the MNRAS server. The first two columns show the coordinates referenced to the 2MASS positions. The following columns list the calibrated magnitudes with their errors and standard error of the calibration offset, and flags for multiple detection and proximity in $J, H$ and $K_s$, respectively. The calibrated magnitudes have not been corrected for interstellar extinction. The multiple detection flag is set to 1 if the source is detected in more than 1 field of view, and 0 otherwise. The proximity flag is set to 1 if there are nearby sources within a radius of 3 arcsec.

In the left panel of Fig.~\ref{colmag1} we show a colour-magnitude diagram of all sources detected in our survey of the SMC. The lower and upper dashed lines indicate the 90\% completeness limit of the moderately populated region (see Table~\ref{comptable}) and the saturation limit, respectively. The numbers of stars in bins of area 0.025$\times$0.025 mag$^2$ were computed and the fiducial colour was applied according to the number density of stars in each bin (see the annotated colour wedge). Because of the limiting magnitude and saturation effects, the lower right area and the upper part of the colour-magnitude diagram are sparsely populated. Note that photometry for the red giants with moderate dust extinction is complete. However, care must be taken for stars with high mass-loss rates whose fluxes are significantly attenuated due to circumstellar extinction.

\begin{table*}
\caption[]{The last five records from the variable source catalogue. Successive columns contain the Mean position, the standard deviation of magnitudes $sd_\lambda$, the Ferreira Lopes variability indices, $I^{(3)^\prime} _{\textrm{pfc}}$ and  $I^{(3)} _{\textrm{fi}}$, the Stetson variability indices ${}^{\lambda_1,\lambda_2}J_{\textrm{WS}}$, the position on the photometric reference image, the calibrated magnitude ${}^{\lambda}m$ and its error ${}^{\lambda}e$, the error in the calibration offset ${}^{\lambda}ec$, the proximity flag, the  name of the time-series data file, and the number of available exposures follows for the $J$, $H$, and $K_s$ bands. The catalogue is in increasing order of Right Ascension. The full version of the catalogue is available on the MNRAS server.}
\label{variable2}
\begin{center}
\begin{tabular}{cccccccccccccccccccccccccccccccccc}
\hline
R.A. & DEC &
${}^{J}sd$ & ${}^{H}sd$ & ${}^{K_s}sd$ & $I^{(3)^\prime} _{\textrm{pfc}}$ & $I^{(3)} _{\textrm{fi}}$ & ${}^{J,H}J_{\textrm{WS}}$ & ${}^{H,K_s}J_{\textrm{WS}}$ & ${}^{J,K_s}J_{\textrm{WS}}$ & 
RA ($J$) & DEC ($J$) & ${}^{J}m$ & ${}^{J}e$ & ${}^{J}ec$ & Proximity & Name of & Number of & 
RA ($H$) & DEC ($H$) & ${}^{H}m$ & ${}^{H}e$ & ${}^{H}ec$ & Proximity & Name of & Number of &
RA ($K_s$) & DEC ($K_s$) & ${}^{K_s}m$ & ${}^{K_s}e$ & ${}^{K_s}ec$ & Proximity & Name of & Number of \\
\multicolumn{2}{c}{[degree]} & 
\multicolumn{3}{c}{[mag]} & \multicolumn{5}{c}{} &
\multicolumn{2}{c}{[degree]} & \multicolumn{3}{c}{[mag]} & flag & time-series data & observations &
\multicolumn{2}{c}{[degree]} & \multicolumn{3}{c}{[mag]} & flag & time-series data & observations &
\multicolumn{2}{c}{[degree]} & \multicolumn{3}{c}{[mag]} & flag & time-series data & observations \\
\hline
15.455597 & -72.816695 & 0.077 & 0.057 & 0.007 & 0.0117 & 0.482 & 6.745 & 1.182 & 1.174 & 15.455584 & -72.816685 & 12.671 & 0.014 & 0.001 & 0 & 15.455597-72.816695.J.dat & 129 & 15.455642 & -72.816713 & 11.677 & 0.011 & 0.001 & 0 & 15.455597-72.816695.H.dat & 128 & 15.455566 & -72.816687 & 10.951 & 0.010 & 0.001 & 0 & 15.455597-72.816695.K.dat & 129 \\
15.456643 & -72.878019 & 0.049 & 0.120 & 0.011 & 0.0048 & 0.482 & 4.245 & 1.022 & 1.267 & 15.456606 & -72.878017 & 12.481 & 0.018 & 0.001 & 0 & 15.456643-72.878019.J.dat & 129 & 15.456708 & -72.878023 & 11.453 & 0.016 & 0.001 & 0 & 15.456643-72.878019.H.dat & 128 & 15.456616 & -72.878016 & 10.983 & 0.012 & 0.001 & 0 & 15.456643-72.878019.K.dat & 129 \\
15.459809 & -72.839262 & 0.060 & -99.999 & 0.090 & -99.9999 & -99.999 & -99.999 & -99.999 & 0.398 & 15.459809 & -72.839266 & 13.853 & 0.011 & 0.001 & 0 & 15.459809-72.839262.J.dat & 129 & 999.999999 & 999.999999 & 99.999 & 99.999 & 99.999 & 0 & ------------------------- & 0 & 15.459808 & -72.839258 & 13.002 & 0.009 & 0.001 & 0 & 15.459809-72.839262.K.dat & 129 \\
15.477461 & -72.972895 & 0.309 & 0.498 & 0.308 & 0.1512 & 0.686 & 26.106 & 18.533 & 13.473 & 15.477449 & -72.972881 & 14.201 & 0.014 & 0.001 & 0 & 15.477461-72.972895.J.dat & 131 & 15.477470 & -72.972895 & 12.960 & 0.012 & 0.001 & 0 & 15.477461-72.972895.H.dat & 130 & 15.477464 & -72.972910 & 11.739 & 0.017 & 0.001 & 0 & 15.477461-72.972895.K.dat & 128 \\
15.509218 & -73.334506 & 0.109 & 0.120 & 0.066 & 0.0034 & 0.263 & 0.788 & 0.314 & 0.420 & 15.509239 & -73.334508 & 15.961 & 0.021 & 0.002 & 0 & 15.509218-73.334506.J.dat & 121 & 15.509207 & -73.334501 & 15.348 & 0.016 & 0.001 & 0 & 15.509218-73.334506.H.dat & 124 & 15.509209 & -73.334508 & 15.259 & 0.027 & 0.001 & 0 & 15.509218-73.334506.K.dat & 120 \\
\hline
\end{tabular}
\end{center}
\end{table*}

\subsection{Variable source catalogue}
A total of 60\,961 variable source candidates were detected. Among them, 1\,063 sources passed the verification process described in the previous section and were identified as real variable sources (See Table~\ref{candidate}). The variable source catalogue lists the coordinates of the variables, the variability indices, the number of available exposures for each variable, and other relevant information. 

\subsubsection{The columns of the variable source catalogue}
The last five records of the variable source catalogue are shown in Table~\ref{variable2}. The table has many columns, but it is repetitive for the $J, H$ and $K_s$ bands from the left to the right. Explanations for each column of the table are given below: \\
\noindent Columns 1-2 : Mean coordinates of the variable source. This coordinate is the mean value if the source is detected in more than one wavebands. \\
\noindent Columns 3-5 : Standard deviation of magnitudes in $J, H$, and $K_s$. If unavailable, ``-99.999" is given.\\
\noindent Column 6-7 : Ferreira Lopes variability indices. If unavailable, ``-99.9999" and ``-99.999" are given for $I^{(3)^\prime} _{\textrm{pfc}}$ and $I^{(3)} _{\textrm{fi}}$ , respectively.\\
\noindent Columns 8-10 : Stetson variability indices for $J\&H$, $H\&K_s$ and $J\&K_s$ pairs, respectively. If unavailable, ``-99.999" is given. \\
\noindent Columns 11-12 : Coordinates of the variable source determined on the photometric reference image. If unavailable, ``999.999999" is given. \\
\noindent Columns 13-15 : Calibrated magnitude, its error and error in the conversion offset. These are the same as the ones in the photometric catalogue. The calibrated magnitude has not been corrected for interstellar extinction. If unavailable, ``99.999" is given.\\
\noindent Column 16 : Proximity flag. If another photometric reference star was present within 3 arcsec of the corresponding position given in Columns 10-11, the flag is set to 1. Otherwise it is set to 0. This is identical to the $m_p$ flag in the point source catalogue \\
\noindent Column 17 : Name of the time-series data file. The name denotes the ``mean coordinate" of the variable source given in Columns 1-2. If unavailable, ``---" is given.\\
\noindent Column 18 : Number of available exposures, $N_\textrm{exp}$, for the variable source. If unavailable, ``0" is given.\\
Columns 11 to 18 are then repeated for each of the $H$ and $K_s$ bands.

\begin{table*}
\caption[]{The first five and last five records from the time-series data for a variable source (14.278202$-$72.827084.H.dat) in the $H$ band. The Heliocentric Julian day, the time-series calibrated magnitude ${}^{\lambda}m$, the error of the time-series magnitude calculated by the differential image analysis $e_\textrm{diff}$, the total systematic error computed from errors in the reference magnitude and calibration offset $e_\textrm{sys}$, and the name of the survey region and field of view in which the variable source detected. The full version of the data is available on the MNRAS server.}
\label{time1}
\begin{center}
\begin{tabular}{ccccc}
\hline
HJD & ${}^{\lambda}m$ & $e_{\textrm{diff}}$ & $e_{\textrm{sys}}$ & Name of \\
$[$day$]$ & \multicolumn{3}{c}{[mag]} & region \& field of view \\
\hline
2452092.615110 & 14.437 & 0.082 & 0.005 & SMC0055$-$7250D \\
2452212.423518 & 13.094 & 0.013 & 0.005 & SMC0055$-$7250D \\
2452213.391569 & 13.201 & 0.021 & 0.005 & SMC0055$-$7250D \\
2452215.292796 & 13.082 & 0.015 & 0.005 & SMC0055$-$7250D \\
2452251.285825 & 12.890 & 0.011 & 0.005 & SMC0055$-$7250D \\
\multicolumn{5}{c}{........} \\
2458077.307696 & 12.617 & 0.007 & 0.005 & SMC0055$-$7250D \\
2458082.336482 & 12.372 & 0.009 & 0.005 & SMC0055$-$7250D \\
2458083.319762 & 12.357 & 0.007 & 0.005 & SMC0055$-$7250D \\
2458092.411123 & 12.478 & 0.009 & 0.005 & SMC0055$-$7250D \\
2458093.327080 & 12.459 & 0.012 & 0.005 & SMC0055$-$7250D \\
\hline
\end{tabular}
\end{center}
\end{table*}

\subsection{Time-series data}
\label{sec:timeseries}
The multi-epoch photometric data (time-series data) of variable sources are also published with this paper. Each variable source has at least one time-series data file for the corresponding waveband in which it is detected. The file name of the time-series data indicates the coordinates and waveband, e.g., ``14.278202$-$72.827084.H.dat". If the source is detected in more than one waveband, the designated coordinate is the mean of the coordinates determined independently for each waveband. The time-series data file contains the information described below.

\subsubsection{The columns of the time-series data catalogue}
An example of the time-series data is shown in Table~\ref{time1}. The data file consists of 5 columns as follows:\\
\noindent Column 1 : Heliocentric Julian day.\\
\noindent Column 2 : Calibrated time-series magnitude. It has not been corrected for interstellar extinction. A value of 99.999 is given for the poor measurement. \\
\noindent Column 3 : Error of the time-series magnitude calculated by the differential image analysis, i.e., error in the first term in the right hand side in equation \ref{henkan}. \\
\noindent Column 4 : Total systematic error determined from errors in the reference magnitude and calibration offset, i.e., error in the term of $m_0^j+\textrm{conversion offset}$ in equation \ref{henkan}. \\
\noindent Column 5 : Name of the survey region and field of view in which the variable source was detected.\\

Note that the number of available exposures is not necessarily the same among the three wavebands. This is due to several reasons, but the main one is (infrequent) readout or reset failures in one or two detectors among the three SIRIUS detectors. In these cases, the corresponding raw data are entirely left out. A value of 99.999 is given for column 2 (and 3) if the exposure data are available but the photometry is of poor quality.

\section{Using the catalogues}
The main objective of this paper is to introduce our survey and publish the data. Detailed discussion of each type of variable sources is beyond the scope of this paper but is in preparation. Here we just demonstrate the catalogue data.

\begin{figure*}
\centering
\includegraphics[scale=0.488,bb=0 0 511 748,angle=0]{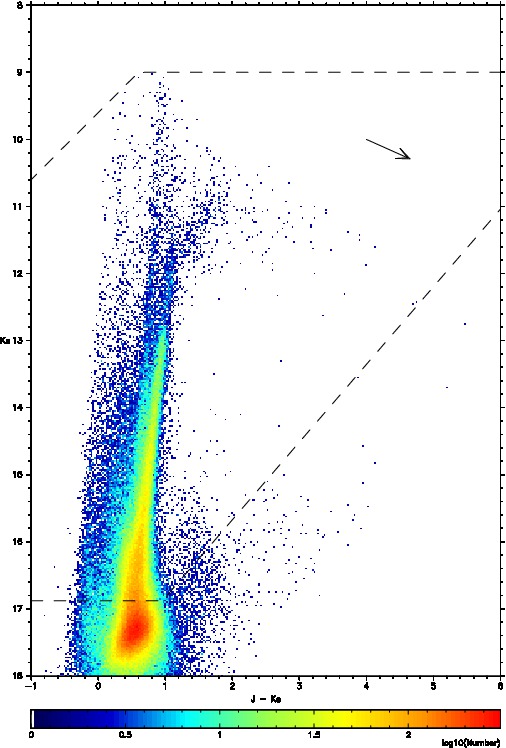}
\raisebox{8mm}{\includegraphics[scale=0.488,bb=10 0 511 748,angle=0]{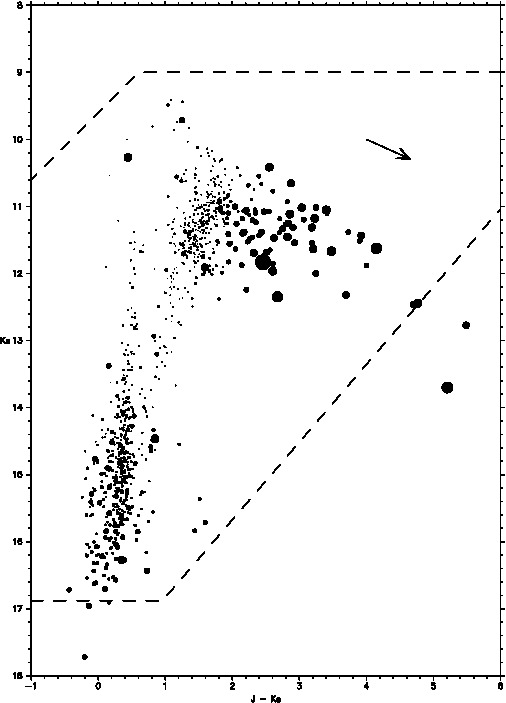}}
\caption{Left panel: Colour-magnitude diagram of all sources detected in the SMC. The bin size is 0.025 $\times$ 0.025 mag. The fiducial colour is applied according to the number density of sources or chance of variability in each bin (see the annotated colour wedge). The reddening vector is drawn for E$_{B-V} = 1$ (note this is much larger than the interstellar reddening for the SMC). The dashed lines indicate the 90 percent completeness limit of the moderately populated field (0055--7310H) given in table \ref{comptable} as a representative value, and the saturation limit, respectively. Right panel: Similar to the left panel, but including only variable sources. The size of the mark is in proportion to the standard deviation of the $J$-band time-series magnitude.}
\label{colmag1}
\end{figure*}

\begin{figure}
\centering
\includegraphics[scale=0.33,bb=0 0 504 720,angle=-90]{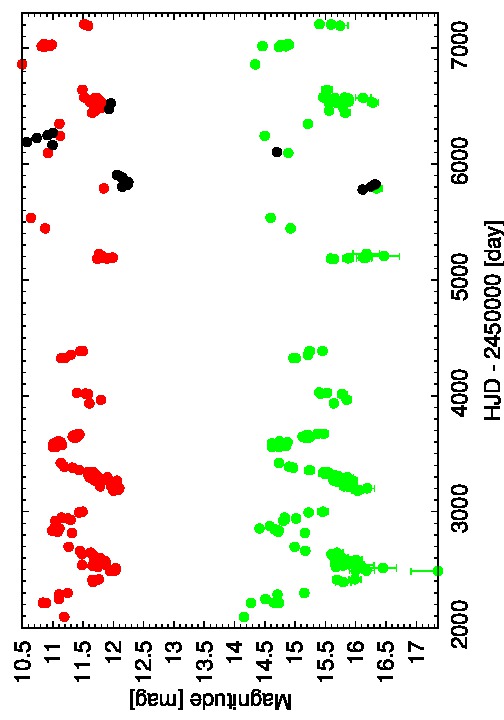}
\vspace{10pt}

\includegraphics[scale=0.33,bb=0 0 504 720,angle=-90]{figure11-1.jpg}
\vspace{10pt}

\includegraphics[scale=0.33,bb=0 0 504 720,angle=-90]{figure11-1.jpg}
\caption{Light curves of three known Mira-like variables in the survey area. Their R.A. and DEC positions in our catalogue are indicated at the top of each panel. For clarity, the $H$ data are not included. Green and Red dots show our time-series data at $J$ and $K_s$, respectively. For comparison, the VMC-DR4 $J$ and $K_s$ data are also shown as black dots. Error bars are shown, but are usually smaller than or comparable to the size of the symbols.}
\label{lightcurve}
\end{figure}

\begin{figure}
\centering
\includegraphics[scale=0.33,bb=0 0 504 720,angle=-90]{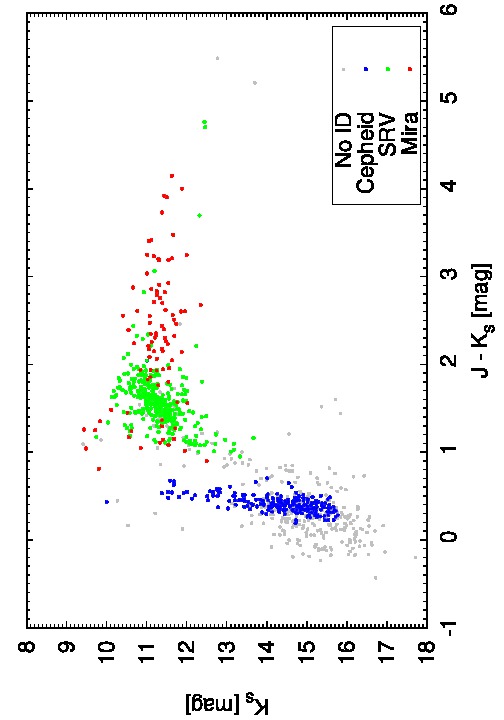}
\includegraphics[scale=0.33,bb=0 0 504 720,angle=-90]{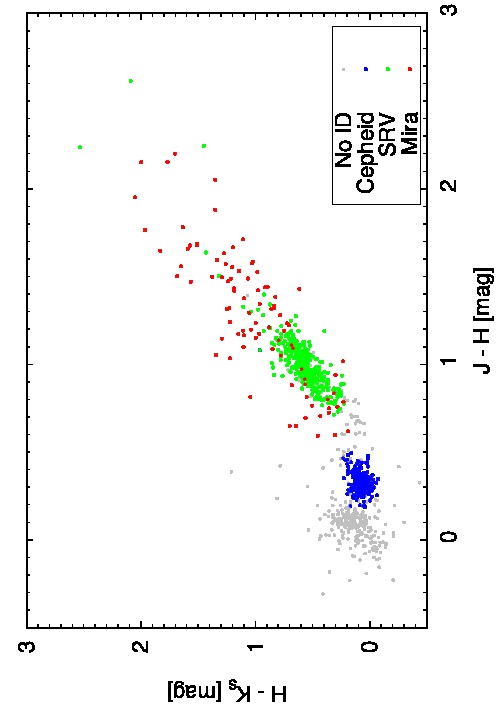}
\caption{Colour-magnitude and colour-colour diagrams of variable sources with OGLE classes.}
\label{colmagcolcol}
\end{figure}

\subsection{Colour magnitude diagram of variable sources}
In the right panel of Fig.~\ref{colmag1}, we show the colour-magnitude diagram of variable sources that passed our evaluation processes. The size of the mark is in proportion to the standard deviation of the $J$-band time-series magnitude, such that the radius of the mark is equal to $1/10$ of the standard deviation. It is obvious that the light variation phenomenon is ubiquitous in the colour-magnitude diagram. It is notable that almost 100\%  of the bright red giants are variable. Many of the luminous and blue sources are also found to be variable. Fig.~\ref{colmag1} only scratches the surface of the available survey data. Many interesting types of variable source are detected and we believe that our data will be useful for a wide variety of research topics.

\subsection{Sample light curves}
As is clear from Fig.~\ref{colmag1}, variable sources with a wide range of colours and luminosities are detected in this survey. Fig.~\ref{lightcurve} provides an illustration of our data for three Mira variables, together with $J$ and $K_s$ observations from VMC-DR4 for the same stars. VMC-DR4 covers a time period when we had rather few measurements. The overall agreement between these data-sets is good, but it would be worth making a detailed analysis of possible differences between the photometric systems before the combined data are used.

\subsection{Comparison to the OGLE survey}
The OGLE project provides huge datasets of variable sources towards the Magellanic Clouds and the data are readily accessible\footnote{OGLE homepage: http://ogledb.astrouw.edu.pl/$^\sim$ogle/}. \citet{soszynski2010} and \citet{soszynski2011} published catalogues of classical Cepheids and long period variables in the SMC. The OGLE survey detected 1\,350 classical Cepheids, 95 Miras and 513 semi-regular variables within the 1$^\circ \times$1$^\circ$ area in common with our survey. We queried the OGLE-III database to extract known variable sources satisfying the following positional criteria: 11.9806 $\le$ RA [degree] $\le$15.5166 and $-$73.3441 $\le$ Decl. [degree] $\le-$72.3215. Among them are 236 classical Cepheids, 94 Miras and 332 semi-regulars that are identified as variables in our survey. Obviously our survey missed very short period Cepheid variables that are faint at near-infrared wavelengths. The sole missing Mira (OGLE-SMC-LPV-06978) is located at the very edge of our survey area. Some of the missing semi-regular variables are located just beyond the edge of our survey area, but most of the others were recognised as variable source candidates, but were eliminated by variability identification processes. It is likely that their amplitudes of light variation are small. Meanwhile there should be some infrared sources that are not detected by OGLE and to test this we selected red sources in an arbitrary manner ($J-H > 1.3$ or $H-K_s > 0.9$) and looked for them in the OGLE database. We found several infrared variables that have no OGLE counterpart. A full discussion of the infrared variables will be made in a separate paper (Ita et al., in preparation). Fig.~\ref{colmagcolcol} shows colour-coded colour-magnitude and two-colour diagrams of variable sources with the above-mentioned three OGLE classes identified in red for Miras, green for semi-regular variables and blue for classical Cepheids, while gray is used for objects without an OGLE classification. 

\section{Conclusions}
A more than fifteen-years near-infrared variable source survey towards the Large and Small Magellanic Clouds has been carried out. This paper publishes the resulting photometric catalogue, a variable source catalogue, and time-series data for the Small Magellanic Cloud. The catalogue provides a moderately deep, multi-epoch survey in the near-infrared covering a large area of the SMC. Further papers in this series will discuss these results.

\section*{Acknowledgments}
Y.I. gratefully acknowledges the hospitality and generous support of Dr. Margaret Meixner and her colleagues during his long term visit to the STScI where this work was written. Y.I is also very grateful to the staffs at the South African Astronomical Observatory (SAAO) for their very helpful support over a decade. We thank Dr. Jungmi Kwon for reading the manuscript and helping us to improve the text. The IRSF project is a collaboration between Nagoya University and the SAAO supported by the Grants-in-Aid for Scientific Research on Priority Areas (A) (No. 10147207 and No. 10147214) and Optical \& Near-Infrared Astronomy Inter-University Cooperation Program, from the Ministry of Education, Culture, Sports, Science and Technology (MEXT) of Japan and the National Research Foundation (NRF) of South Africa. This work is supported by the Grant-in-Aid for Encouragement of Young Scientists (B) No.21740142 from the MEXT of Japan. This work is also supported by the Brain Circulation Program (R2301) by Japan Society for the Promotion of Science. MWF, PAW and JWM gratefully acknowledge the receipt of research grants from the NRF of South Africa. This paper uses observations made at the SAAO. This publication makes use of data products from the Two Micron All Sky Survey, which is a joint project of the University of Massachusetts and the Infrared Processing and Analysis Center/California Institute of Technology, funded by the National Aeronautics and Space Administration and the National Science Foundation. This paper also makes use of the OGLE database and the SIMBAD database of the CDS.








\appendix

\section[]{The central coordinates of the fields observed in the SMC}
\begin{center}
\begin{table}
\caption{The central coordinates of observed fields of view in the SMC}
\label{surveyregionsmc}
\begin{tabular}{ccrr}
\hline
SMC0050-7250 & A & 00$^\textrm{h}$ 52$^\textrm{m}$ 04.45$^\textrm{s}$ & -72$^{\circ}$ 57$^{\prime}$ 00.00$^{\prime\prime}$ \\
             & B & 00$^\textrm{h}$ 50$^\textrm{m}$ 28.95$^\textrm{s}$ & -72$^{\circ}$ 57$^{\prime}$ 00.00$^{\prime\prime}$ \\
             & C & 00$^\textrm{h}$ 48$^\textrm{m}$ 53.45$^\textrm{s}$ & -72$^{\circ}$ 57$^{\prime}$ 00.00$^{\prime\prime}$ \\
             & D & 00$^\textrm{h}$ 52$^\textrm{m}$ 03.81$^\textrm{s}$ & -72$^{\circ}$ 50$^{\prime}$ 00.00$^{\prime\prime}$ \\
             & E & 00$^\textrm{h}$ 50$^\textrm{m}$ 28.95$^\textrm{s}$ & -72$^{\circ}$ 50$^{\prime}$ 00.00$^{\prime\prime}$ \\
             & F & 00$^\textrm{h}$ 48$^\textrm{m}$ 54.09$^\textrm{s}$ & -72$^{\circ}$ 50$^{\prime}$ 00.00$^{\prime\prime}$ \\
             & G & 00$^\textrm{h}$ 52$^\textrm{m}$ 03.19$^\textrm{s}$ & -72$^{\circ}$ 43$^{\prime}$ 00.00$^{\prime\prime}$ \\
             & H & 00$^\textrm{h}$ 50$^\textrm{m}$ 28.95$^\textrm{s}$ & -72$^{\circ}$ 43$^{\prime}$ 00.00$^{\prime\prime}$ \\
             & I & 00$^\textrm{h}$ 48$^\textrm{m}$ 54.71$^\textrm{s}$ & -72$^{\circ}$ 43$^{\prime}$ 00.00$^{\prime\prime}$ \\
SMC0050-7310 & A & 00$^\textrm{h}$ 52$^\textrm{m}$ 01.09$^\textrm{s}$ & -73$^{\circ}$ 17$^{\prime}$ 00.00$^{\prime\prime}$ \\
             & B & 00$^\textrm{h}$ 50$^\textrm{m}$ 23.74$^\textrm{s}$ & -73$^{\circ}$ 17$^{\prime}$ 00.00$^{\prime\prime}$ \\
             & C & 00$^\textrm{h}$ 48$^\textrm{m}$ 46.40$^\textrm{s}$ & -73$^{\circ}$ 17$^{\prime}$ 00.00$^{\prime\prime}$ \\
             & D & 00$^\textrm{h}$ 52$^\textrm{m}$ 00.43$^\textrm{s}$ & -73$^{\circ}$ 10$^{\prime}$ 00.00$^{\prime\prime}$ \\
             & E & 00$^\textrm{h}$ 50$^\textrm{m}$ 23.74$^\textrm{s}$ & -73$^{\circ}$ 10$^{\prime}$ 00.00$^{\prime\prime}$ \\
             & F & 00$^\textrm{h}$ 48$^\textrm{m}$ 47.05$^\textrm{s}$ & -73$^{\circ}$ 10$^{\prime}$ 00.00$^{\prime\prime}$ \\
             & G & 00$^\textrm{h}$ 51$^\textrm{m}$ 59.79$^\textrm{s}$ & -73$^{\circ}$ 03$^{\prime}$ 00.00$^{\prime\prime}$ \\
             & H & 00$^\textrm{h}$ 50$^\textrm{m}$ 23.74$^\textrm{s}$ & -73$^{\circ}$ 03$^{\prime}$ 00.00$^{\prime\prime}$ \\
             & I & 00$^\textrm{h}$ 48$^\textrm{m}$ 47.70$^\textrm{s}$ & -73$^{\circ}$ 03$^{\prime}$ 00.00$^{\prime\prime}$ \\
SMC0051-7230 & A & 00$^\textrm{h}$ 52$^\textrm{m}$ 07.67$^\textrm{s}$ & -72$^{\circ}$ 37$^{\prime}$ 00.00$^{\prime\prime}$ \\
             & B & 00$^\textrm{h}$ 50$^\textrm{m}$ 33.95$^\textrm{s}$ & -72$^{\circ}$ 37$^{\prime}$ 00.00$^{\prime\prime}$ \\
             & C & 00$^\textrm{h}$ 49$^\textrm{m}$ 00.23$^\textrm{s}$ & -72$^{\circ}$ 37$^{\prime}$ 00.00$^{\prime\prime}$ \\
             & D & 00$^\textrm{h}$ 52$^\textrm{m}$ 07.07$^\textrm{s}$ & -72$^{\circ}$ 30$^{\prime}$ 00.00$^{\prime\prime}$ \\
             & E & 00$^\textrm{h}$ 50$^\textrm{m}$ 33.95$^\textrm{s}$ & -72$^{\circ}$ 30$^{\prime}$ 00.00$^{\prime\prime}$ \\
             & F & 00$^\textrm{h}$ 49$^\textrm{m}$ 00.84$^\textrm{s}$ & -72$^{\circ}$ 30$^{\prime}$ 00.00$^{\prime\prime}$ \\
             & G & 00$^\textrm{h}$ 52$^\textrm{m}$ 06.47$^\textrm{s}$ & -72$^{\circ}$ 23$^{\prime}$ 00.00$^{\prime\prime}$ \\
             & H & 00$^\textrm{h}$ 50$^\textrm{m}$ 33.95$^\textrm{s}$ & -72$^{\circ}$ 23$^{\prime}$ 00.00$^{\prime\prime}$ \\
             & I & 00$^\textrm{h}$ 49$^\textrm{m}$ 01.44$^\textrm{s}$ & -72$^{\circ}$ 23$^{\prime}$ 00.00$^{\prime\prime}$ \\
SMC0055-7230 & A & 00$^\textrm{h}$ 56$^\textrm{m}$ 33.71$^\textrm{s}$ & -72$^{\circ}$ 37$^{\prime}$ 00.00$^{\prime\prime}$ \\
             & B & 00$^\textrm{h}$ 55$^\textrm{m}$ 00.00$^\textrm{s}$ & -72$^{\circ}$ 37$^{\prime}$ 00.00$^{\prime\prime}$ \\
             & C & 00$^\textrm{h}$ 53$^\textrm{m}$ 26.28$^\textrm{s}$ & -72$^{\circ}$ 37$^{\prime}$ 00.00$^{\prime\prime}$ \\
             & D & 00$^\textrm{h}$ 56$^\textrm{m}$ 33.11$^\textrm{s}$ & -72$^{\circ}$ 30$^{\prime}$ 00.00$^{\prime\prime}$ \\
             & E & 00$^\textrm{h}$ 55$^\textrm{m}$ 00.00$^\textrm{s}$ & -72$^{\circ}$ 30$^{\prime}$ 00.00$^{\prime\prime}$ \\
             & F & 00$^\textrm{h}$ 53$^\textrm{m}$ 26.88$^\textrm{s}$ & -72$^{\circ}$ 30$^{\prime}$ 00.00$^{\prime\prime}$ \\
             & G & 00$^\textrm{h}$ 56$^\textrm{m}$ 32.51$^\textrm{s}$ & -72$^{\circ}$ 23$^{\prime}$ 00.00$^{\prime\prime}$ \\
             & H & 00$^\textrm{h}$ 55$^\textrm{m}$ 00.00$^\textrm{s}$ & -72$^{\circ}$ 23$^{\prime}$ 00.00$^{\prime\prime}$ \\
             & I & 00$^\textrm{h}$ 53$^\textrm{m}$ 27.48$^\textrm{s}$ & -72$^{\circ}$ 23$^{\prime}$ 00.00$^{\prime\prime}$ \\
SMC0055-7250 & A & 00$^\textrm{h}$ 56$^\textrm{m}$ 35.49$^\textrm{s}$ & -72$^{\circ}$ 57$^{\prime}$ 00.00$^{\prime\prime}$ \\
             & B & 00$^\textrm{h}$ 55$^\textrm{m}$ 00.00$^\textrm{s}$ & -72$^{\circ}$ 57$^{\prime}$ 00.00$^{\prime\prime}$ \\
             & C & 00$^\textrm{h}$ 53$^\textrm{m}$ 24.50$^\textrm{s}$ & -72$^{\circ}$ 57$^{\prime}$ 00.00$^{\prime\prime}$ \\
             & D & 00$^\textrm{h}$ 56$^\textrm{m}$ 34.86$^\textrm{s}$ & -72$^{\circ}$ 50$^{\prime}$ 00.00$^{\prime\prime}$ \\
             & E & 00$^\textrm{h}$ 55$^\textrm{m}$ 00.00$^\textrm{s}$ & -72$^{\circ}$ 50$^{\prime}$ 00.00$^{\prime\prime}$ \\
             & F & 00$^\textrm{h}$ 53$^\textrm{m}$ 25.13$^\textrm{s}$ & -72$^{\circ}$ 50$^{\prime}$ 00.00$^{\prime\prime}$ \\
             & G & 00$^\textrm{h}$ 56$^\textrm{m}$ 34.24$^\textrm{s}$ & -72$^{\circ}$ 43$^{\prime}$ 00.00$^{\prime\prime}$ \\
             & H & 00$^\textrm{h}$ 55$^\textrm{m}$ 00.00$^\textrm{s}$ & -72$^{\circ}$ 43$^{\prime}$ 00.00$^{\prime\prime}$ \\
             & I & 00$^\textrm{h}$ 53$^\textrm{m}$ 25.75$^\textrm{s}$ & -72$^{\circ}$ 43$^{\prime}$ 00.00$^{\prime\prime}$ \\
SMC0055-7310 & A & 00$^\textrm{h}$ 56$^\textrm{m}$ 37.34$^\textrm{s}$ & -73$^{\circ}$ 17$^{\prime}$ 00.00$^{\prime\prime}$ \\
             & B & 00$^\textrm{h}$ 55$^\textrm{m}$ 00.00$^\textrm{s}$ & -73$^{\circ}$ 17$^{\prime}$ 00.00$^{\prime\prime}$ \\
             & C & 00$^\textrm{h}$ 53$^\textrm{m}$ 22.65$^\textrm{s}$ & -73$^{\circ}$ 17$^{\prime}$ 00.00$^{\prime\prime}$ \\
             & D & 00$^\textrm{h}$ 56$^\textrm{m}$ 36.69$^\textrm{s}$ & -73$^{\circ}$ 10$^{\prime}$ 00.00$^{\prime\prime}$ \\
             & E & 00$^\textrm{h}$ 55$^\textrm{m}$ 00.00$^\textrm{s}$ & -73$^{\circ}$ 10$^{\prime}$ 00.00$^{\prime\prime}$ \\
             & F & 00$^\textrm{h}$ 53$^\textrm{m}$ 23.31$^\textrm{s}$ & -73$^{\circ}$ 10$^{\prime}$ 00.00$^{\prime\prime}$ \\
             & G & 00$^\textrm{h}$ 56$^\textrm{m}$ 36.04$^\textrm{s}$ & -73$^{\circ}$ 03$^{\prime}$ 00.00$^{\prime\prime}$ \\
             & H & 00$^\textrm{h}$ 55$^\textrm{m}$ 00.00$^\textrm{s}$ & -73$^{\circ}$ 03$^{\prime}$ 00.00$^{\prime\prime}$ \\
             & I & 00$^\textrm{h}$ 53$^\textrm{m}$ 23.95$^\textrm{s}$ & -73$^{\circ}$ 03$^{\prime}$ 00.00$^{\prime\prime}$ \\
\end{tabular}
\end{table}
\end{center}

\begin{center}
\begin{table}
\begin{tabular}{ccrr}
SMC0059-7230 & A & 01$^\textrm{h}$ 00$^\textrm{m}$ 59.76$^\textrm{s}$ & -72$^{\circ}$ 37$^{\prime}$ 00.00$^{\prime\prime}$ \\
             & B & 00$^\textrm{h}$ 59$^\textrm{m}$ 26.04$^\textrm{s}$ & -72$^{\circ}$ 37$^{\prime}$ 00.00$^{\prime\prime}$ \\
             & C & 00$^\textrm{h}$ 57$^\textrm{m}$ 52.32$^\textrm{s}$ & -72$^{\circ}$ 37$^{\prime}$ 00.00$^{\prime\prime}$ \\
             & D & 01$^\textrm{h}$ 00$^\textrm{m}$ 59.15$^\textrm{s}$ & -72$^{\circ}$ 30$^{\prime}$ 00.00$^{\prime\prime}$ \\
             & E & 00$^\textrm{h}$ 59$^\textrm{m}$ 26.04$^\textrm{s}$ & -72$^{\circ}$ 30$^{\prime}$ 00.00$^{\prime\prime}$ \\
             & F & 00$^\textrm{h}$ 57$^\textrm{m}$ 52.92$^\textrm{s}$ & -72$^{\circ}$ 30$^{\prime}$ 00.00$^{\prime\prime}$ \\
             & G & 01$^\textrm{h}$ 00$^\textrm{m}$ 58.55$^\textrm{s}$ & -72$^{\circ}$ 23$^{\prime}$ 00.00$^{\prime\prime}$ \\
             & H & 00$^\textrm{h}$ 59$^\textrm{m}$ 26.04$^\textrm{s}$ & -72$^{\circ}$ 23$^{\prime}$ 00.00$^{\prime\prime}$ \\
             & I & 00$^\textrm{h}$ 57$^\textrm{m}$ 53.52$^\textrm{s}$ & -72$^{\circ}$ 23$^{\prime}$ 00.00$^{\prime\prime}$ \\
SMC0100-7250 & A & 01$^\textrm{h}$ 01$^\textrm{m}$ 06.54$^\textrm{s}$ & -72$^{\circ}$ 57$^{\prime}$ 00.00$^{\prime\prime}$ \\
             & B & 00$^\textrm{h}$ 59$^\textrm{m}$ 31.04$^\textrm{s}$ & -72$^{\circ}$ 57$^{\prime}$ 00.00$^{\prime\prime}$ \\
             & C & 00$^\textrm{h}$ 57$^\textrm{m}$ 55.55$^\textrm{s}$ & -72$^{\circ}$ 57$^{\prime}$ 00.00$^{\prime\prime}$ \\
             & D & 01$^\textrm{h}$ 01$^\textrm{m}$ 05.91$^\textrm{s}$ & -72$^{\circ}$ 50$^{\prime}$ 00.00$^{\prime\prime}$ \\
             & E & 00$^\textrm{h}$ 59$^\textrm{m}$ 31.04$^\textrm{s}$ & -72$^{\circ}$ 50$^{\prime}$ 00.00$^{\prime\prime}$ \\
             & F & 00$^\textrm{h}$ 57$^\textrm{m}$ 56.18$^\textrm{s}$ & -72$^{\circ}$ 50$^{\prime}$ 00.00$^{\prime\prime}$ \\
             & G & 01$^\textrm{h}$ 01$^\textrm{m}$ 05.29$^\textrm{s}$ & -72$^{\circ}$ 43$^{\prime}$ 00.00$^{\prime\prime}$ \\
             & H & 00$^\textrm{h}$ 59$^\textrm{m}$ 31.04$^\textrm{s}$ & -72$^{\circ}$ 43$^{\prime}$ 00.00$^{\prime\prime}$ \\
             & I & 00$^\textrm{h}$ 57$^\textrm{m}$ 56.80$^\textrm{s}$ & -72$^{\circ}$ 43$^{\prime}$ 00.00$^{\prime\prime}$ \\
SMC0100-7310 & A & 01$^\textrm{h}$ 01$^\textrm{m}$ 13.59$^\textrm{s}$ & -73$^{\circ}$ 17$^{\prime}$ 00.00$^{\prime\prime}$ \\
             & B & 00$^\textrm{h}$ 59$^\textrm{m}$ 36.25$^\textrm{s}$ & -73$^{\circ}$ 17$^{\prime}$ 00.00$^{\prime\prime}$ \\
             & C & 00$^\textrm{h}$ 57$^\textrm{m}$ 58.91$^\textrm{s}$ & -73$^{\circ}$ 17$^{\prime}$ 00.00$^{\prime\prime}$ \\
             & D & 01$^\textrm{h}$ 01$^\textrm{m}$ 12.94$^\textrm{s}$ & -73$^{\circ}$ 10$^{\prime}$ 00.00$^{\prime\prime}$ \\
             & E & 00$^\textrm{h}$ 59$^\textrm{m}$ 36.25$^\textrm{s}$ & -73$^{\circ}$ 10$^{\prime}$ 00.00$^{\prime\prime}$ \\
             & F & 00$^\textrm{h}$ 57$^\textrm{m}$ 59.56$^\textrm{s}$ & -73$^{\circ}$ 10$^{\prime}$ 00.00$^{\prime\prime}$ \\
             & G & 01$^\textrm{h}$ 01$^\textrm{m}$ 12.29$^\textrm{s}$ & -73$^{\circ}$ 03$^{\prime}$ 00.00$^{\prime\prime}$ \\
             & H & 00$^\textrm{h}$ 59$^\textrm{m}$ 36.25$^\textrm{s}$ & -73$^{\circ}$ 03$^{\prime}$ 00.00$^{\prime\prime}$ \\
             & I & 00$^\textrm{h}$ 58$^\textrm{m}$ 00.21$^\textrm{s}$ & -73$^{\circ}$ 03$^{\prime}$ 00.00$^{\prime\prime}$ \\
\hline
\hline
\end{tabular}
\end{table}
\end{center}


\bsp	
\label{lastpage}
\end{document}